\documentstyle[12pt,amsfonts,amsmath,amscd]{article}

\begin{document}

\pagestyle{empty}
\begin{flushright}
KYUSHU-HET-72
\end{flushright}
\vspace*{2cm}

\vspace{0.3cm}
\begin{center}
\LARGE{\textbf{Non-Supersymmetric \\
$T^2/Z_N$ and $T^4/Z_N$ Orbifolds}}\\[1cm]
\renewcommand{\thefootnote}{\fnsymbol{footnote}}
{\large Kentaro Kawazu}\footnote{kawazu@higgs.phys.kyushu-u.ac.jp} 
\\[-1mm]
{\normalsize Department of Physics, Kyushu University, Fukuoka 812-8581, Japan 
}\\
[10mm]
\renewcommand{\thefootnote}{\arabic{footnote}}

\small{\textbf{Abstract}} \\[3mm]
\end{center}

\begin{center}
\begin{minipage}[h]{15.5cm}
\small{
We compute the perturbative tachyonic and massless spectra
of Type II and Type 0 string theories on 
non-supersymmetric $T^2/Z_N$ orbifolds, and those on $T^4/Z_N$ ones. 
Comparing the spectra with one another, 
we obtain insight about the degeneracy of states 
and find several pairs of Type 0 orbifolds which could be identified
with each other.
}
\end{minipage}
\end{center}

\newpage

\setcounter{page}{1} \pagestyle{plain}
\renewcommand{\thefootnote}{\arabic{footnote}}
\setcounter{footnote}{0}

\section{Introduction}

There have been many studies of supersymmetric both
compact and non-compact orbifolds \cite{bsf,hep-th/0005153};
however, there are many mathematically consistent 
non-supersymmetric orbifolds \cite{nso}. 
In recent years, non-compact non-supersymmetric orbifolds have 
been studied intensively \cite{tac},    
yet compact ones have not been paid much attention to;
therefore we should advance studies of these.
In this paper, we address only Type II and Type 0 string theories, 
and thus regrettably do not deal with heterotic ones.

As well known, we can regard Type 0B/0A string theory 
as the Type IIB/IIA orbifold
with the spacetime fermion number operator $(-1)^{F_S}$ \cite{dhsw},
but in this paper we treat Type 0 string theories 
as distinct from Type II ones,
for $(-1)^{F_S}$ acts trivially in spacetime.
The spectra of Type II and Type 0 string theories are as follows \cite{pol}:
\begin{equation}
\begin{array}{lccccr}
\textrm{IIB} & : & (\textrm{NS+,NS+})&(\textrm{R+,NS+})&(\textrm{NS+,R+}) 
&(\textrm{R+,R+}),\\
\textrm{IIA} & : & (\textrm{NS+,NS+})&(\textrm{R+,NS+})&(\textrm{NS+,R}-)
&(\textrm{R+,R}-),\\
\textrm{0B} & : & (\textrm{NS+,NS+})&(\textrm{NS}-,\textrm{NS}-)&
(\textrm{R+,R+})&(\textrm{R}-,\textrm{R}-),\\
\textrm{0A} & : & (\textrm{NS+,NS+})&(\textrm{NS}-,\textrm{NS}-)&
(\textrm{R+,R}-)&(\textrm{R}-,\textrm{R}+).
\end{array}
\label{baba}
\end{equation}
Of course, 
we can consider supersymmetric orbifolds only of Type II strings.
   
Although each spectrum of Type II and Type 0 string theories
on a non-supersymmetric orbifold has tachyons, 
which cause vacuum instability, we expect that
tachyon condensation scenarios rescue these from this difficulty \cite{tac}.
Also for absence of fermions in Type 0 orbifolds,
we anticipate that D-branes are the key to solve it,
for there appear fermionic modes of 
open strings stretching between D-branes 
in a certain configuration \cite{hep-th/9906055}.  
Taking these into consideration,
it is worthwhile to study non-supersymmetric orbifolds also.

For example, $Z_2$ orbifolds of Type II and Type 0 strings have been
discussed with addressing the relation between 
the D-brane spectra and the  
corresponding K-theories \cite{hepth9910109}.
It is natural that one desire to analyze higher order orbifolds
in the same manner.
Therefore, in this paper, for the first step of the analysis,
we compute and enumerate the perturbative tachyonic and massless spectra of 
non-supersymmetric $T^2/Z_N$ and $T^4/Z_N$ orbifolds
of Type II and Type 0 strings.
In doing that, compactifications are useful to give restrictions
on the orders of orbifolds \cite{dhvw,hepth9207111}.
We shall also enumerate the massless spectra of Type II string theories 
on a supersymmetric $T^4/Z_N$ orbifold for comparison.

This paper is organized as follows.
In section 2 we introduce the basic notation.
We enumerate the tachyonic and massless spectra of Type II and Type 0 string 
theories on a $T^2/Z_N$ orbifold in section 3,
and those on a $T^4/Z_N$ in section 4.
In section 5 we extract some rules of the spectra
from the results in section 4.
Summary and discussion are given in the final section.
We have included an appendix to transform the spectra in section 3
to those on a $T^2$$\times$$T^2/Z_N$ orbifold. 

\section{The basic notation}

First, we define the complex linear combinations
\begin{equation}
\begin{array}{l}
Z^i=\frac{1}{\sqrt{2}}(X^{2i}+iX^{2i+1}) , \\[5pt]
\overline{Z^i}=\frac{1}{\sqrt{2}}(X^{2i}-iX^{2i+1}) ,
\hspace{3mm}i=2,3,4.
\end{array}
\end{equation}
We can always choose the axes so that the rotation is of the form
\begin{equation}
\theta=\textrm{exp}[2\pi i(v_2J_{45}+v_3J_{67}+v_4J_{89})],
\label{rotation}
\end{equation}
which acts on $Z^i$ as $\theta Z^i$=$e^{2\pi iv_i}Z^i$. 
Here, $v$=$(v_2,v_3,v_4)$, or $(0,v_2,v_3,v_4)$,
denotes the twist vector corresponding to the generator $\theta$,
and in this paper always $v_2=0$.
The $v_i$ is of the form $k/N$, for $\theta^N$=$1$.

The twist vector $v$ of the generator $\theta$ has several restrictions.
The restriction from geometry is that
$\theta$ must act crystallographically on the torus lattice.
All such twists have been obtained in \cite{dhvw,hepth9207111}.
Modular invariance imposes the constraint
\begin{equation}
N\sum_i v_i=0 \hspace{1mm}\textrm{mod}\hspace{1mm}2.
\label{mod}
\end{equation} 
Supersymmetry gives the additional condition
\begin{equation}
\pm v_2 \pm v_3 \pm v_4=0\hspace{1mm}\textrm{mod}\hspace{1mm}2. 
\label{susy}
\end{equation}
We can obtain non-supersymmetric orbifolds by relaxing (\ref{susy}).
Then, we can classify the orbifolds into two types:
one does not include the (0,0,1)=$(-1)^{F_S}$ twist 
and the other does. 
Type II string theories on the former 
have fermionic states in the spectra.
Even if the former satisfies condition (\ref{susy}),
we can consider Type 0 string theories on that 
and also those are bosonic ones.
For the former,
the order of a Type 0 orbifold is the same as that of a Type II,
as long as the generator is the same.
For the latter, the orders depend on whether we consider 
Type II orbifold or Type 0 one; however, 
we can identify Type 0B/0A string theory on $Z_N$ orbifold with
Type IIB/IIA one on $Z_{2N}$ orbifold,
provided the generator is the same.  
This comes from Type 0 string theories 
being regarded as the Type II orbifolds with $(-1)^{F_S}$.
In this paper, if a orbifold includes the $(-1)^{F_S}$ twist,
we consider only Type 0 string theories, not Type II ones.

In section 3 and 4 we list the generators of $T^2/Z_{N}$
and $T^4/Z_{N}$ orbifolds, respectively.
Then, following the way in the appendix of \cite{hepth0202057},
we compute and enumerate the spectra of Type II and Type 0 string 
theories on those.   


\section{$T^2/Z_N$ orbifolds}
\label{t2}

In this section we concentrate our attention on $T^2/Z_N$ orbifolds
of Type II and Type 0 strings. 
We obtain six twist vectors, which are listed in Table \ref{tab1},
compatible with the toroidal compactification $T^2$.

\begin{table}[htb]
\begin{center}
 \begin{tabular}{|c|c|c|} 
 \hline
  $(v_2,v_3,v_4)$ & $Z_N$ (for Type II) & $Z_N$ (for Type 0)\\ 
 \hline
 (0,0,1) & $Z_2$ & 1 \\
 \hline
 (0,0,$\frac{2}{3}$) & $Z_3$ & $Z_3$ \\
 \hline
 (0,0,$\frac{1}{2}$) & $Z_4$ & $Z_2$ \\
 \hline
 (0,0,$\frac{1}{3}$) & $Z_6$ & $Z_3$ \\
 \hline
 (0,0,$\frac{1}{4}$) & $Z_8$ & $Z_4$ \\
 \hline
 (0,0,$\frac{1}{6}$) & $Z_{12}$ & $Z_6$ \\
 \hline
 \end{tabular}
\end{center}
\caption{Twist vectors with one non-zero component.}
\label{tab1}
\end{table}


As mentioned in the previous section, 
we identify $v$=(0,0,1) with the spacetime 
fermion number operator $(-1)^{F_S}$, 
thus it is the identical operator for Type 0 string theories.
We compute the tachyonic and massless
spectra of Type II and Type 0 string theories on  
other $T^2/Z_N$ orbifolds
under $SO$(8) $\to$ $SO$(6)$\times$$SO$(2),
with $\mathbf{8_v}$=($\mathbf{6}$,0)+($\mathbf{1}$,1)+($\mathbf{1}$,$-1$)
and 
$\mathbf{8_s}$=($\mathbf{4}$,$-\frac{1}{2}$)+($\overline{\mathbf{4}}$,$\frac{1}{2}$). 
In each $T^2/Z_N$ orbifold, 
there always appear from the untwisted NS-NS sector
$\mathbf{20}$+$\mathbf{15}$+$\mathbf{1}$, 
which correspond to graviton, antisymmetric tensor and dilaton.
Therefore we omit these states in the spectra.
In addition, all the $T^2/Z_N$ orbifolds of Type II and Type 0 strings
are non-supersymmetric,
hence there always appear tachyons in the spectra.
We represent those as $\mathbf{1_{m^2}}$,
where the subscript indicates the value of $m^2_R$=$m^2_L$.


As the first example, we consider Type II and Type 0 string theories
on the $T^2/Z_3$ orbifold with $v$=(0,0,$\frac{2}{3}$).
Among $T^2/Z_N$ orbifolds, 
only this orbifold does not include the $(-1)^{F_S}$=(0,0,1) twist.
Therefore, fermionic states appear 
in the spectra of Type II string theoris on this orbifold.
In fact, we find the tachyonic and massless spectrum of Type IIB string 
theory is
\begin{equation}
\begin{array}{lcl}
\theta^0 & : & (\mathbf{1}+\mathbf{4}+
\mathbf{4}+\mathbf{1}+\mathbf{15})+(\mathbf{1}+
\overline{\mathbf{4}}+\overline{\mathbf{4}}+\mathbf{1}+\mathbf{15}),\\
\theta & : & 3(\mathbf{1_{-\frac{1}{3}}})
+\textrm{3}(\mathbf{1}+\mathbf{4}+
\mathbf{4}+\mathbf{1}+\mathbf{15}), \\
\theta^2 & : & \textrm{3}(\mathbf{1_{-\frac{1}{3}}})
+\textrm{3}(\mathbf{1}+
\overline{\mathbf{4}}+\overline{\mathbf{4}}+\mathbf{1}+\mathbf{15}),
\end{array}
\end{equation}  
where $\theta^0$ denotes the untwisted sector,
each $\theta^n$ ($n\ne0$) does $n$th twisted sector.  
Similarly, we find that of Type IIA string theory is
\begin{equation}
\begin{array}{lcl}
\theta^0 & : & \textrm{2}(\mathbf{1}+\mathbf{4}+
\overline{\mathbf{4}}+\mathbf{6}+\mathbf{10}),\\
\theta & : & \textrm{3}(\mathbf{1_{-\frac{1}{3}}})
+\textrm{3}(\mathbf{1}+\mathbf{4}+
\overline{\mathbf{4}}+\mathbf{6}+\mathbf{10}), \\
\theta^2 & : & \textrm{3}(\mathbf{1_{-\frac{1}{3}}})
+\textrm{3}(\mathbf{1}+
\mathbf{4}+\overline{\mathbf{4}}+\mathbf{6}+\mathbf{10}).
\end{array}
\end{equation} 
As mentioned in section 2, 
we also consider Type 0 string theories on this orbifold.
The tachyonic and massless spectrum of Type 0B string theory is
\begin{equation}
\begin{array}{lcl}
\theta^0 & : & \mathbf{1_{-\frac{1}{2}}}+
\textrm{2}(\mathbf{1})+
\textrm{2}(\mathbf{1}+\mathbf{15}+\mathbf{1}+\mathbf{15}),\\
\theta,\theta^2 & : & 
6(\mathbf{1_{-\frac{1}{3}}}+\mathbf{1_{-\frac{1}{6}}}+\mathbf{1})+
\textrm{6}(\mathbf{1}+\mathbf{15}+\mathbf{1}+\mathbf{15}).
\end{array}
\end{equation}  
And that of Type 0A string theory is
\begin{equation}
\begin{array}{lcl}
\theta^0 & : & \mathbf{1_{-\frac{1}{2}}}+
\textrm{2}(\mathbf{1})
+\textrm{2}(\mathbf{6}+\mathbf{10}+\mathbf{6}+\mathbf{10}),\\
\theta,\theta^2 & : & 6(\mathbf{1_{-\frac{1}{3}}}+\mathbf{1_{-\frac{1}{6}}}+
\mathbf{1})+\textrm{6}(\mathbf{6}+\mathbf{10}+\mathbf{6}+\mathbf{10}). 
\end{array}
\end{equation}
As we expected, Type 0 string theories have doubled R-R states;  
$\mathbf{1}$+$\mathbf{15}$+$\mathbf{1}$+$\mathbf{15}$ in Type 0B 
and $\mathbf{6}$+$\mathbf{10}$+$\mathbf{6}$+$\mathbf{10}$ in Type 0A
are twice as many as those in Type IIB and IIA respectively.


The other $T^2/Z_N$ orbifolds include the $(-1)^{F_S}$ twist,
therefore, we consider only Type 0 string theories. 
On the $T^2/Z_2$ orbifold with $v$=(0,0,$\frac{1}{2}$), 
the tachyonic and massless spectrum of Type 0B string theory is
\begin{equation}
\begin{array}{lcl}
\theta^0 & : & \mathbf{1_{-\frac{1}{2}}}+\textrm{4}(\mathbf{1})
+\textrm{2}(\mathbf{1}+\mathbf{15}+\mathbf{1}+\mathbf{15}),\\
\theta & : & 8(\mathbf{1_{-\frac{1}{4}}})
+\textrm{4}(\mathbf{1}+\mathbf{15}+\mathbf{1}+\mathbf{15}),
\end{array}
\end{equation} 
and that of Type 0A string theory is
\begin{equation}
\begin{array}{lcl}
\theta^0 & : & \mathbf{1_{-\frac{1}{2}}}
+\textrm{4}(\mathbf{1})+
\textrm{2}(\mathbf{6}+\mathbf{10}+\mathbf{6}+\mathbf{10}),\\
\theta & : & 8(\mathbf{1_{-\frac{1}{4}}})
+\textrm{4}(\mathbf{6}+\mathbf{10}+\mathbf{6}+\mathbf{10}). 
\end{array}
\end{equation}
The feature of these spectra   
is four scalars $\mathbf{1}$ in each untwisted sector;
these are doubled compared to those of any other $T^2/Z_N$ orbifolds
of Type 0 strings.
This results from $\frac{1}{2}$ being half-integer.


Next, we consider Type 0 string theories on the $T^2/Z_3$ orbifold
with $v$=(0,0,$\frac{1}{3}$).
The tachyonic and massless spectrum 
of Type 0B string theory on this orbifold is   
\begin{equation}
\begin{array}{lcl}
\theta^0 & : & \mathbf{1_{-\frac{1}{2}}}+\textrm{2}(\mathbf{1})+
\textrm{2}(\mathbf{1}+\mathbf{15}+\mathbf{1}+\mathbf{15}),\\
\theta,\theta^2 & : & 
6(\mathbf{1_{-\frac{1}{3}}}+\mathbf{1_{-\frac{1}{6}}}+\mathbf{1})+
\textrm{6}(\mathbf{1}+\mathbf{15}+\mathbf{1}+\mathbf{15}). \\
\end{array}
\end{equation} 
We also find that of Type 0A string theory is
\begin{equation}
\begin{array}{lcl}
\theta^0 & : & \mathbf{1_{-\frac{1}{2}}}+\textrm{2}(\mathbf{1})+
\textrm{2}(\mathbf{6}+\mathbf{10}+\mathbf{6}+\mathbf{10}),\\
\theta,\theta^2 & : & 
6(\mathbf{1_{-\frac{1}{3}}}+\mathbf{1_{-\frac{1}{6}}}+\mathbf{1})+
\textrm{6}(\mathbf{6}+\mathbf{10}+\mathbf{6}+\mathbf{10}). \\
\end{array}
\end{equation}
We notice that these spectra are the same as those on 
the $T^2/Z_3$ orbifold with $v$=(0,0,$\frac{2}{3}$) 
of Type 0B and Type 0A strings, respectively.
Therefore, we can identify the two $T^2/Z_3$ orbifolds 
of Type 0 strings, one with $v$=(0,0,$\frac{2}{3}$) and the
other with $v$=(0,0,$\frac{1}{3}$), with each other at 
tachyonic and massless level at least.


The two following $T^2/Z_N$ orbifolds include a few distinct twists 
from those of generators. 
Let us take Type 0 string theories on the $T^2/Z_4$ orbifold with
$v$=(0,0,$\frac{1}{4}$).
We find the tachyonic and massless spectrum of Type 0B string theory
on this orbifold is  
\begin{equation}
\begin{array}{lcl}
\theta^0 & : & \mathbf{1_{-\frac{1}{2}}}+\textrm{2}(\mathbf{1})
+\textrm{2}(\mathbf{1}+\mathbf{15}+\mathbf{1}+\mathbf{15}),\\
\theta,\theta^3 & : & 4(\mathbf{1_{-\frac{3}{8}}})+
\textrm{8}(\mathbf{1_{-\frac{1}{8}}})
+\textrm{4}(\mathbf{1}+\mathbf{15}+\mathbf{1}+\mathbf{15}), \\
\theta^2 & : & 6(\mathbf{1_{-\frac{1}{4}}})
+\textrm{3}(\mathbf{1}+\mathbf{15}+\mathbf{1}+\mathbf{15}).
\end{array}
\end{equation} 
We notice that the number ratio of the states in the $\theta^2$ sector
is the same as that in the $\theta$ sector of the $T^2/Z_2$ orbifold,
for two twisted sectors have the same twist vector (0,0,$\frac{1}{2}$).
However, the numbers of the states, the degeneracy of the states,
are different. This difference depends on whether the twist vector 
(0,0,$\frac{1}{2}$) corresponds to the generator or the subgroup.
We can also notice the same relation in Type 0A string theory on
this orbifold: we find the tachyonic and massless spectrum of it is  
\begin{equation}
\begin{array}{lcl}
\theta^0 & : & \mathbf{1_{-\frac{1}{2}}}+\textrm{2}(\mathbf{1})+
\textrm{2}(\mathbf{6}+\mathbf{10}+\mathbf{6}+\mathbf{10}),\\
\theta,\theta^3 & : & 4(\mathbf{1_{-\frac{3}{8}}})
+\textrm{8}(\mathbf{1_{-\frac{1}{8}}})
+\textrm{4}(\mathbf{6}+\mathbf{10}+\mathbf{6}+\mathbf{10}), \\
\theta^2 & : & 6(\mathbf{1_{-\frac{1}{4}}})
+\textrm{3}(\mathbf{6}+\mathbf{10}+\mathbf{6}+\mathbf{10}).
\end{array}
\end{equation}


As the final $T^2/Z_N$ orbifold, 
we consider Type 0 string theories on the $T^2/Z_6$ orbifold
with $v$=(0,0,$\frac{1}{6}$).
We find the tachyonic and massless spectrum of Type 0B string theory 
on this orbifold is   
\begin{equation}
\begin{array}{lcl}
\theta^0 & : & \mathbf{1_{-\frac{1}{2}}}+\textrm{2}(\mathbf{1})
+\textrm{2}(\mathbf{1}+\mathbf{15}+\mathbf{1}+\mathbf{15}),\\
\theta,\theta^5 & : & \textrm{2}(\mathbf{1_{-\frac{5}{12}}})
+\textrm{2}(\mathbf{1_{-\frac{1}{4}}})
+\textrm{4}(\mathbf{1_{-\frac{1}{12}}})
+\textrm{2}(\mathbf{1}+\mathbf{15}+\mathbf{1}+\mathbf{15}), \\
\theta^2,\theta^4 & : & 4(\mathbf{1_{-\frac{1}{3}}}
+\mathbf{1_{-\frac{1}{6}}}+\mathbf{1})
+\textrm{4}(\mathbf{1}+\mathbf{15}+\mathbf{1}+\mathbf{15}),\\
\theta^3 & : & 4(\mathbf{1_{-\frac{1}{4}}})
+\textrm{2}(\mathbf{1}+\mathbf{15}+\mathbf{1}+\mathbf{15}).
\end{array}
\end{equation}
In this case, the number ratio of the states in the $\theta^2$+$\theta^4$
sector is the same as that in the $\theta$ sector of the $T^2/Z_3$ orbifold
with $v$=(0,0,$\frac{1}{3}$).
In addition, the ratio in the $\theta^3$ sector is the same as that in 
the $\theta$ sector of the $T^2/Z_2$ orbifold with $v$=(0,0,$\frac{1}{2}$).
Of course, these relations are also true for Type 0A string theory on
this orbifold. The tachyonic and massless spectrum of it is 
\begin{equation}
\begin{array}{lcl}
\theta^0 & : & \mathbf{1_{-\frac{1}{2}}}+\textrm{2}(\mathbf{1})
+\textrm{2}(\mathbf{6}+\mathbf{10}+\mathbf{6}+\mathbf{10}),\\
\theta,\theta^5 & : & \textrm{2}(\mathbf{1_{-\frac{5}{12}}})+
\textrm{2}(\mathbf{1_{-\frac{1}{4}}})
+\textrm{4}(\mathbf{1_{-\frac{1}{12}}})
+\textrm{2}(\mathbf{6}+\mathbf{10}+\mathbf{6}+\mathbf{10}), \\
\theta^2,\theta^4 & : & 4(\mathbf{1_{-\frac{1}{3}}}
+\mathbf{1_{-\frac{1}{6}}}+\mathbf{1})
+\textrm{4}(\mathbf{6}+\mathbf{10}+\mathbf{6}+\mathbf{10}),\\
\theta^3 & : & 4(\mathbf{1_{-\frac{1}{4}}})
+\textrm{2}(\mathbf{6}+\mathbf{10}+\mathbf{6}+\mathbf{10}).
\end{array}
\end{equation}

As we have seen,  
in the $T^2/Z_N$ orbifolds of Type 0 strings,  
the number ratios of the states are the same 
among the twisted sectors with the same twist vector.
In the next section we shall observe that 
the ratios are not preserved among several twisted sectors
of Type II and Type 0 string theories on $T^4/Z_N$ orbifolds,
even if those have the same twist vector. 
We transform the spectra obtained in this section 
into $T^2$$\times$$T^2/Z_N$ orbifolds version in the appendix 
in order to compare these with those. 

\section{$T^4/Z_N$ orbifolds}
\label{t4}
This section consists of two subsections.
We focus on $T^4/Z_N$ orbifolds not including the $(-1)^{F_S}$
twist and those doing in subsection \ref{t4.1} and \ref{t4.2},
respectively.

We compute the spectra of Type II and Type 0 string theories on
a $T^4/Z_N$ orbifold under 
$SO$(4) $\simeq$ $SU$(2)$\times$$SU$(2) besides
$SO$(8) $\to$ $SO$(4)$\times$$SO$(4).
Therefore, the vector ($\underline{\pm1,0}$), 
the spinor ($\underline{\frac{1}{2},-\frac{1}{2}}$),
and the spinor $\pm(\frac{1}{2},\frac{1}{2})$ become
the $(\frac{1}{2},\frac{1}{2})$, $(\frac{1}{2},0)$ and
$(0,\frac{1}{2})$ representations respectively, 
where underlines mean permutations.   
In addition, we represent tachyons as $(0,0)_{m^2}$,
where the subscript indicates the value of $m^2_R$=$m^2_L$
as in section \ref{t2}.

\subsection{Orbifolds not including the $(-1)^{F_S}$ twist}
\label{t4.1}

In this subsection we focus on orbifolds not including the
$(-1)^{F_S}$ twist, listed in Table \ref{tab2}.
As mentioned in section 2,
the order of a Type 0 orbifold is the same as that of a Type II,
as long as the generator is the same. 

\begin{table}[htb]
\begin{center}
 \begin{tabular}{|c|c||c|c||c|c|} 
 \hline
  $(v_2,v_3,v_4)$ & $Z_N$ & $(v_2,v_3,v_4)$ & $Z_N$ & $(v_2,v_3,v_4)$ & $Z_N$\\ 
 \hline
 (0,$\frac{1}{2}$,$\frac{1}{2}$) & $Z_2^{\ast}$ & 
 (0,$\frac{1}{5}$,$\frac{3}{5}$) & $Z_5$ &
 (0,$\frac{1}{8}$,$\frac{3}{8}$) & $Z_8$ \\
 \hline
 (0,$\frac{1}{3}$,$\frac{1}{3}$) & $Z_3^{\ast}$ &
 (0,$\frac{1}{6}$,$\frac{1}{6}$) & $Z_6^{\ast}$ &
 (0,$\frac{1}{10}$,$\frac{3}{10}$) & $Z_{10}$\\
 \cline{1-2} 
 \cline{5-6}
 (0,$\frac{1}{4}$,$\frac{1}{4}$) & $Z_4^{\ast}$ &
 (0,$\frac{1}{6}$,$\frac{3}{6}$)=(0,$\frac{1}{6}$,$\frac{1}{2}$) & $Z_6$ &
 (0,$\frac{1}{12}$,$\frac{5}{12}$) & $Z_{12}$\\
 (0,$\frac{1}{4}$,$\frac{3}{4}$) & $Z_4$ & 
 (0,$\frac{1}{6}$,$\frac{5}{6}$) & $Z_6$ &
 (0,$\frac{1}{12}$,$\frac{7}{12}$) & $Z_{12}$ \\
 \hline
\end{tabular}
\end{center}
\caption{twist vectors not including the $(-1)^{F_S}$ twist
with two non-zero components.
We take absolute value of two components of a twist vector.
The star $\ast$ on a $Z_N$ means that 
Type II string theories on that orbifold are supersymmetric.}
\label{tab2}
\end{table}

As well known, the massless spectrum of a
supersymmetric Type II orbifold forms some 
supersymmetric multiplets.
On any supersymmetric $T^4/Z_N$ orbifolds, 
the states in the untwisted sector of Type IIB string theory
form one supergravity multiplet 
\begin{equation}
(1,1)+4(1,\frac{1}{2})+5(0,0)
\end{equation} 
and some tensor multiplets 
\begin{equation}
(0,1)+4(0,\frac{1}{2})+5(0,0).
\label{tensor}
\end{equation}
Several tensor multiplets (\ref{tensor}) 
also appear in the twisted sectors: thus
we regard the number ratio of the states in (\ref{tensor})
as the standard one in the twisted sectors with a supersymmetric
twist vector of Type IIB string theory on a $T^4/Z_N$ orbifold.

Similarly, 
the states in the untwisted sector of Type IIA string theory 
on a $T^4/Z_N$ supersymmetric orbifold form one supergravity multiplet
\begin{equation}
(1,1)+(1,0)+(0,1)+(0,0)+4(\frac{1}{2},\frac{1}{2}) 
+2(1,\frac{1}{2})+2(\frac{1}{2},1)+2(0,\frac{1}{2})+2(\frac{1}{2},0)
\end{equation}
and some vector multiplets 
\begin{equation}
(\frac{1}{2},\frac{1}{2})+2(0,\frac{1}{2})+2(\frac{1}{2},0)+4(0,0),
\label{vector}
\end{equation}
and the states in the twisted sectors 
form several vector multiplets (\ref{vector}).
We also regard the number ratio of the states in (\ref{vector}) 
as the standard one in the twisted sectors with a supersymmetric 
twist vector of Type IIA string theory on a $T^4/Z_N$ orbifold.


As the first example of $T^4/Z_N$ orbifolds,
let us take the $T^4/Z_2$ orbifold with $v$=(0,$\frac{1}{2}$,$-\frac{1}{2}$). 
As well known, Type II string theories on this orbifold are
supersymmetric, therefore we find the supersymmetric massless spectrum
of Type IIB string theory on this orbifold is 
\begin{equation}
\begin{array}{lcl}
\theta^0 & : & \textrm{(1 supergravity multiplet)+(5 tensor multiplets)}, \\
\theta & : & \textrm{16 tensor multiplets},
\end{array}
\end{equation}
and that of Type IIA string theory is
\begin{equation}
\begin{array}{lcl}
\theta^0 & : & \textrm{(1 supergravity multiplet)+(4 vector multiplets)},\\
\theta & : & \textrm{16 vector multiplets}.
\end{array}
\end{equation}

In addition, we can also consider Type 0 string theories on this orbifold.
In General, the untwisted sectors of Type 0 string theories on
a $T^4/Z_N$ orbifold always include
\begin{equation}
(0,0)_{-\frac{1}{2}}+(1,1)+(1,0)+(0,1)+(0,0):
\end{equation}
thus we omit these states except for the tachyon $(0,0)_{-\frac{1}{2}}$ 
from their spectra. 
That is, we omit graviton, antisymmetric tensor and dilaton. 
The remaining tachyonic and massless states of Type 0B string theory 
on this $T^4/Z_2$ orbifold are
\begin{equation}
\begin{array}{lcl}
\theta^0 & : & (0,0)_{-\frac{1}{2}}+8(1,0)+8(0,1)+32(0,0),  \\
\theta & : & 16(1,0)+16(0,1)+160(0,0),
\end{array}
\end{equation}
and those of Type 0A string theory are
\begin{equation}
\begin{array}{lcl}
\theta^0 & : & (0,0)_{-\frac{1}{2}}+16(\frac{1}{2},\frac{1}{2})+16(0,0),  \\
\theta & : & 32(\frac{1}{2},\frac{1}{2})+128(0,0).
\end{array}
\end{equation}
We regard the number ratio of the states in the $\theta$ sector 
of Type 0B/0A string theory 
as the standard one in the twisted sectors with the twist vector
(0,$\frac{1}{2}$,$-\frac{1}{2}$) of Type 0B/0A. 


Next, we consider Type II and Type 0 string theories on  
the $T^4/Z_3$ orbifold with $v$=(0,$\frac{1}{3}$,$-\frac{1}{3}$). 
As on the previous orbifold, Type II string theories on it are supersymmetric.
Hence, we find the supersymmetric massless spectrum of Type IIB 
string theory is
\begin{equation}
\begin{array}{lcl}
\theta^0 & : & \textrm{(1 supergravity multiplet)+(3 tensor multiplets)}, \\
\theta,\theta^2 & : & \textrm{18 tensor multiplets}.
\end{array}
\end{equation}
We notice the total of multiplets on this $T^4/Z_3$ orbifold 
is the same as that on the previous $T^4/Z_2$ orbifold. 
The same thing also appear in Type IIA string theory as follows: 
\begin{equation}
\begin{array}{lcl}
\theta^0 & : & \textrm{(1 supergravity multiplet)+(2 vector multiplets)}, \\
\theta,\theta^2 & : & \textrm{18 vector multiplets}.
\end{array}
\end{equation}

In Type 0 string theories, 
several tachyons appear in both the untwisted sectors
and the twisted sectors.
The tachyonic and massless spectrum of Type 0B 
string theory on this $T^4/Z_3$ orbifold is
\begin{equation}
\begin{array}{lcl}
\theta^0 & : & (0,0)_{-\frac{1}{2}}+6(1,0)+6(0,1)+20(0,0),  \\
\theta,\theta^2 & : & 18(0,0)_{-\frac{1}{6}}+18(1,0)+18(0,1)+108(0,0).
\end{array}
\end{equation}
Similarly, we find that of Type 0A string theory is
\begin{equation}
\begin{array}{lcl}
\theta^0 & : & (0,0)_{-\frac{1}{2}}+12(\frac{1}{2},\frac{1}{2})+8(0,0), \\
\theta,\theta^2 & : & 18(0,0)_{-\frac{1}{6}}
+36(\frac{1}{2},\frac{1}{2})+72(0,0).
\end{array}
\end{equation}
We regard the number ratio of the states, including the tachyons,
in the $\theta$+$\theta^2$ sector of Type 0B/0A string theory 
as the standard one in the twisted sectors  
with the twist vector (0,$\frac{1}{3}$,$-\frac{1}{3}$) of Type 0B/0A. 


The $T^4/Z_4$ orbifold with $v$=(0,$\frac{1}{4}$,$-\frac{1}{4}$) 
is the first example of $T^4/Z_N$ orbifolds 
which have distinct twists from that of the generator.
In fact, we find the supersymmetric massless spectrum of Type IIB 
string theory on this orbifold is
\begin{equation}
\begin{array}{lcl}
\theta^0 & : & \textrm{(1 supergravity multiplet)+(3 tensor multiplets)}, \\
\theta,\theta^3 & : & \textrm{8 tensor multiplets}, \\
\theta^2 & : & \textrm{10 tensor multiplets}.
\end{array}
\end{equation}
Although this orbifold has two different twists,
the total of supersymmetric multiplets on this orbifold is 
the same as that on a previous supersymmetric orbifolds. 
The same thing appear in the supersymmetric massless spectrum of Type IIA
string theory on this orbifold as follows:
\begin{equation}
\begin{array}{lcl}
\theta^0 & : & \textrm{(1 supergravity multiplet)+(2 vector multiplets)}, \\
\theta,\theta^3 & : & \textrm{8 vector multiplets}, \\
\theta^2 & : & \textrm{10 vector multiplets}.
\end{array}
\end{equation}

We also consider Type 0 string theories on this $T^4/Z_4$ orbifold. 
The tachyonic and massless spectrum of Type 0B string theory is
\begin{equation}
\begin{array}{lcl}
\theta^0 & : & (0,0)_{-\frac{1}{2}}+6(1,0)+6(0,1)+20(0,0),  \\
\theta,\theta^3 & : & 8(0,0)_{-\frac{1}{4}}+8(1,0)+8(0,1)+80(0,0), \\
\theta^2 & : & 10(1,0)+10(0,1)+100(0,0).
\end{array}
\end{equation}
We notice easily that the number ratio of the states 
in the $\theta^2$ sector is the same as 
the standard one, i.e. that in the $\theta$ sector of the $T^4/Z_2$ 
orbifold with $v$=(0,$\frac{1}{2}$,$-\frac{1}{2}$).
We can also find the same thing in the tachyonic and massless
spectrum of Type 0A string theory as follows: 
\begin{equation}
\begin{array}{lcl}
\theta^0 & : & (0,0)_{-\frac{1}{2}}+12(\frac{1}{2},\frac{1}{2})+8(0,0), \\
\theta,\theta^3 & : & 8(0,0)_{-\frac{1}{4}}+
16(\frac{1}{2},\frac{1}{2})+64(0,0), \\
\theta^2 & : & 20(\frac{1}{2},\frac{1}{2})+80(0,0).
\end{array}
\end{equation}
We regard the number ratio of the states
in the $\theta+\theta^3$ sector of Type 0B/0A string theory 
as the standard one in the twisted sectors with the
twist vector (0,$\frac{1}{4}$,$-\frac{1}{4}$) of Type 0B/0A.


The $T^4/Z_4$ orbifold with $v$=(0,$\frac{1}{4}$,$\frac{3}{4}$) is  
the first example of non-supersymmetric $T^4/Z_N$ orbifolds.
Therefore, the untwisted sectors of Type II string theories
on this orbifold cannot form a supergravity multiplet.
However, there always appear in the untwisted sectors
\begin{equation}
(1,1)+(1,0)+(0,1)+(0,0),
\end{equation}
which we omit also in the spectra of Type II string theories on
other non-supersymmetric $T^4/Z_4$ orbifolds. 
And also, tachyons appear in the spectra, 
even if we consider Type II string theories.
In fact, the tachyonic and massless spectrum of Type IIB string theory
on this $T^4/Z_4$ orbifold is
\begin{equation}
\begin{array}{lcl}
\theta^0 & : & 4(1,0)+2(0,1)+8(0,\frac{1}{2})+14(0,0),\\
\theta,\theta^3 & : & 8(0,0)_{-\frac{1}{4}}
+8(1,0)+32(\frac{1}{2},0)+40(0,0), \\ 
\theta^2 & : & 10(0,1)+24(0,\frac{1}{2})+50(0,0).
\end{array}
\end{equation}
As indicated in \cite{hepth0202057},
there appear 8 tensor multiplets and
extra 8 tachyons in the $\theta+\theta^3$ sector.
We also notice that the $\theta^2$ sector wants 
$16(0,\frac{1}{2})$ to fill the 10 tensor multiplets.
Similarly, states appearing and disappearing are found in
the tachyonic and massless spectrum of  
Type IIA string theory on this orbifold as follows:
\begin{equation}
\begin{array}{lcl}
\theta^0 & : & 6(\frac{1}{2},\frac{1}{2})+4(\frac{1}{2},0)
+4(0,\frac{1}{2})+8(0,0), \\
\theta,\theta^3 & : & 8(0,0)_{-\frac{1}{4}}+8(\frac{1}{2},\frac{1}{2})
+16(\frac{1}{2},0)+16(0,\frac{1}{2})+32(0,0), \\
\theta^2 & : & 10(\frac{1}{2},\frac{1}{2})
+12(\frac{1}{2},0)+12(0,\frac{1}{2})+40(0,0).
\end{array}
\end{equation}
The $\theta^2$ sector wants
$8(\frac{1}{2},0)+8(0,\frac{1}{2})$ to fill the 10 vector multiplets.
We regard the number ratio of the states in the $\theta$+$\theta^3$
sector of Type IIB/IIA string theory as the standard one 
in the twisted sectors 
with the twist vector (0,$\frac{1}{4}$,$\frac{3}{4}$) of Type IIB/IIA.  

In addition, we consider Type 0 string theories on this $T^4/Z_4$ orbifold.
The tachyonic and massless spectrum of Type 0B string theory on this 
orbifold is
\begin{equation}
\begin{array}{lcl}
\theta^0 & : & (0,0)_{-\frac{1}{2}}+6(1,0)+6(0,1)+20(0,0),  \\
\theta,\theta^3 & : & 8(0,0)_{-\frac{1}{4}}+8(1,0)+8(0,1)+80(0,0), \\
\theta^2 & : & 10(1,0)+10(0,1)+100(0,0).
\end{array}
\end{equation}
And also, that of Type 0A string theory is
\begin{equation}
\begin{array}{lcl}
\theta^0 & : & (0,0)_{-\frac{1}{2}}+12(\frac{1}{2},\frac{1}{2})+8(0,0), \\
\theta,\theta^3 & : & 8(0,0)_{-\frac{1}{4}}+
16(\frac{1}{2},\frac{1}{2})+64(0,0), \\
\theta^2 & : & 20(\frac{1}{2},\frac{1}{2})+80(0,0). 
\end{array}
\end{equation}
While we notice that the tachyonic and massless 
spectrum of Type IIB/IIA string theory on the $T^4/Z_4$ orbifold with
$v$=(0,$\frac{1}{4}$,$\frac{3}{4}$) is different from that with
$v$=(0,$\frac{1}{4}$,$-\frac{1}{4}$),
we obtain the same spectrum of Type 0B/0A string theory on
each $T^4/Z_4$ orbifold.
Therefore the $T^4/Z_4$ orbifold with $v$=(0,$\frac{1}{4}$,$\frac{3}{4}$) is
equivalent to that with $v$=(0,$\frac{1}{4}$,$-\frac{1}{4}$) 
in Type 0 string theories at tachyonic and massless level at least.
On a parallel with Type II string theories,
we regard the number ratio of the states in the $\theta$+$\theta^3$
sector of Type 0B/0A string theory as the standard one in the twisted sectors 
with the twist vector (0,$\frac{1}{4}$,$\frac{3}{4}$) of Type 0B/0A.  


The orbifold with $v$=(0,$\frac{1}{5}$,$\frac{3}{5}$) is  
a non-supersymmetric $Z_5$ orbifold. 
As indicated in \cite{hepth0202057},
the tachyonic and massless spectrum of Type IIB string theory on this
orbifold is
\begin{equation}
\begin{array}{lcl}
\theta^0 & : & 2(1,0)+2(0,1)+4(0,\frac{1}{2})+4(\frac{1}{2},0)+8(0,0), \\
\theta,\theta^4 & : & 10(0,0)_{-\frac{1}{5}}
+10(1,0)+20(\frac{1}{2},0)+20(0,0), \\ 
\theta^2,\theta^3 & : & 10(0,0)_{-\frac{1}{5}}
+10(0,1)+20(0,\frac{1}{2})+20(0,0),
\end{array}
\end{equation}
and that of Type IIA string theory is 
\begin{equation}
\begin{array}{lcl}
\theta^0 & : & 4(\frac{1}{2},\frac{1}{2})+4(\frac{1}{2},0)
+4(0,\frac{1}{2})+4(0,0), \\
\theta,\theta^4 & : & 10(0,0)_{-\frac{1}{5}}+10(\frac{1}{2},\frac{1}{2})
+10(\frac{1}{2},0)+10(0,\frac{1}{2})+10(0,0), \\
\theta^2,\theta^3 & : & 10(0,0)_{-\frac{1}{5}}+10(\frac{1}{2},\frac{1}{2})
+10(\frac{1}{2},0)+10(0,\frac{1}{2})+10(0,0).
\end{array}
\end{equation}
In addition to Type II string theories, we consider Type 0 string theories.
We find the tachyonic and massless spectrum of Type 0B string theory
on this $T^4/Z_5$ orbifold is 
\begin{equation}
\begin{array}{lcl}
\theta^0 & : & (0,0)_{-\frac{1}{2}}+4(1,0)+4(0,1)+12(0,0),  \\
\theta,\theta^4 & : & 10(0,0)_{-\frac{1}{5}}+10(0,0)_{-\frac{1}{10}}+
10(1,0)+10(0,1)+30(0,0), \\
\theta^2,\theta^3 & : & 10(0,0)_{-\frac{1}{5}}+10(0,0)_{-\frac{1}{10}}+
10(1,0)+10(0,1)+30(0,0),
\end{array}
\end{equation}
and that of Type 0A string theory is
\begin{equation}
\begin{array}{lcl}
\theta^0 & : & (0,0)_{-\frac{1}{2}}+8(\frac{1}{2},\frac{1}{2})+4(0,0), \\
\theta,\theta^4 & : & 10(0,0)_{-\frac{1}{5}}+10(0,0)_{-\frac{1}{10}}+
20(\frac{1}{2},\frac{1}{2})+10(0,0), \\
\theta^2,\theta^3 & : & 10(0,0)_{-\frac{1}{5}}+10(0,0)_{-\frac{1}{10}}+
20(\frac{1}{2},\frac{1}{2})+10(0,0).
\end{array}
\end{equation}
The discrete rotation $Z_5$ has no subgroup, 
therefore, in each string theory,
we can regard the number ratio of the states in each
twisted sector as the standard one. 
  
 
The $T^4/Z_6$ orbifold with $v$=(0,$\frac{1}{6}$,$-\frac{1}{6}$) has
the most highest order of all the supersymmetric $T^4/Z_N$ orbifolds. 
Thus, we find the supersymmetric massless spectrum of Type IIB 
string theory on this orbifold is
\begin{equation}
\begin{array}{lcl}
\theta^0 & : & \textrm{(1 supergravity multiplet)+(3 tensor multiplets)}, \\
\theta,\theta^5 & : & \textrm{2 tensor multiplets}, \\
\theta^2,\theta^4 & : & \textrm{10 tensor multiplets}, \\
\theta^3 & : & \textrm{6 tensor multiplets}.
\end{array}
\end{equation}
Although this orbifold has the three different twists,
the total of the supersymmetric multiplets,
one supergravity multiplet and 21 tensor multiplets,
is the same as that on
other supersymmetric $T^4/Z_N$ orbifolds.  
Similar thing also appear in that of Type IIA string theory as follows:
\begin{equation}
\begin{array}{lcl}
\theta^0 & : & \textrm{(1 supergravity multiplet)+(2 vector multiplets)}, \\
\theta,\theta^5 & : & \textrm{2 vector multiplets}, \\
\theta^2,\theta^4 & : & \textrm{10 vector multiplets}, \\
\theta^3 & : & \textrm{6 vector multiplets}. 
\end{array}
\end{equation}
That is, we find again 
single supergravity multiplet and 20 vector multiplets.  

We also consider Type 0 string theories. 
The tachyonic and massless spectrum of Type 0B string theory on
this $T^4/Z_6$ orbifold is
\begin{equation}
\begin{array}{lcl}
\theta^0 & : & (0,0)_{-\frac{1}{2}}+6(1,0)+6(0,1)+20(0,0),  \\
\theta,\theta^5 & : & 2(0,0)_{-\frac{1}{3}}+8(0,0)_{-\frac{1}{6}}+
2(1,0)+2(0,1)+30(0,0), \\
\theta^2,\theta^4 & : & 10(0,0)_{-\frac{1}{6}}
+10(1,0)+10(0,1)+60(0,0), \\
\theta^3 & : & 6(1,0)+6(0,1)+60(0,0).
\end{array}
\end{equation}
We notice the number ratio of the states in the $\theta^2$+$\theta^4$ 
sector is the standard one, i.e. that in the $\theta$ sector 
of the $T^4/Z_3$ orbifold with $v$=(0,$\frac{1}{3}$,$-\frac{1}{3}$).
And also, we notice 
that in the $\theta^3$ sector is the standard one, i.e.
that in the $\theta$ sector 
of the $T^4/Z_2$ orbifold with $v$=(0,$\frac{1}{2}$,$-\frac{1}{2}$) 
Of course, we can confirm this preservation of the ratios
also in Type 0A string theory as follows: 
the tachyonic and massless spectrum of Type 0A string theory on this
orbifold is 
\begin{equation}
\begin{array}{lcl}
\theta^0 & : & (0,0)_{-\frac{1}{2}}+12(\frac{1}{2},\frac{1}{2})+8(0,0), \\
\theta,\theta^5 & : & 2(0,0)_{-\frac{1}{3}}+8(0,0)_{-\frac{1}{6}}+
4(\frac{1}{2},\frac{1}{2})+26(0,0), \\
\theta^2,\theta^4 & : & 10(0,0)_{-\frac{1}{6}}
+20(\frac{1}{2},\frac{1}{2})+40(0,0), \\
\theta^3 & : &  12(\frac{1}{2},\frac{1}{2})+48(0,0).
\end{array}
\end{equation}
While we find the same total massless spectrum of Type IIB/IIA string theory
on any supersymmetric $T^4/Z_N$ orbifolds, 
we find different that of Type 0B/0A string theory on each 
supersymmetric $T^4/Z_N$ orbifold.
This results from the number of the NS-NS states changing,
therefore the total number of the R-R states is 
constant on each supersymmetric orbifold. 
Of course, the spectra of
Type 0 string theories on a \textit{supersymmetric orbifold} 
are not supersymmetric,
but it means the twist vector $v$ corresponding to the generator of 
a orbifold satisfies the condition (\ref{susy}). 

There is no supersymmetric orbifold in the following;
however, this orbifold seems 
the most interesting of all the orbifolds treated in this paper.
That is the $T^4/Z_6$ orbifold with 
$v$=(0,$\frac{1}{6}$,$\frac{3}{6}$)=(0,$\frac{1}{6}$,$\frac{1}{2}$).  
We find the tachyonic and massless spectrum of Type IIB string theory
on this orbifold is 
\begin{equation}
\begin{array}{lcl}
\theta^0 & : & 2(1,0)+2(0,1)+4(0,\frac{1}{2})+10(0,0), \\
\theta,\theta^5 & : & 8(0,0)_{-\frac{1}{6}}+
8(1,0)+16(\frac{1}{2},0)+16(0,0), \\
\theta^2,\theta^4 & : & 4(0,0)_{-\frac{1}{3}}+4(1,0)+4(0,1)
+4(\frac{1}{2},\frac{1}{2})+4(\frac{1}{2},0)+8(0,\frac{1}{2})+12(0,0), \\
\theta^3 & : &  8(0,1)+16(0,\frac{1}{2})+32(0,0).
\end{array}
\end{equation}
By comparison with the standard ratio, i.e. the $\theta$ sector of
the $T^2$$\times$$T^2/Z_3$ orbifold with $v$=(0,0,$\frac{2}{3}$),
the $\theta^2$+$\theta^4$ sector lacks 
4($\frac{1}{2}$,$\frac{1}{2}$)+4($\frac{1}{2}$,0).
We should remark that disappearing 4($\frac{1}{2}$,$\frac{1}{2}$) 
come from the R-R sectors.
This suggests that the corresponding fractional D-branes are also absent.
Although 3$v$ is supersymmetric, the $\theta^3$ sector 
lacks 16(0,$\frac{1}{2}$)+8(0,0) to fill 8 tensor multiplets.
Similarly, we find the tachyonic and massless spectrum
of Type IIA string theory is 
\begin{equation}
\begin{array}{lcl}
\theta^0 & : & 4(\frac{1}{2},\frac{1}{2})+2(\frac{1}{2},0)
+2(0,\frac{1}{2})+6(0,0), \\
\theta,\theta^5 & : & 8(0,0)_{-\frac{1}{6}}+8(\frac{1}{2},\frac{1}{2})+
8(\frac{1}{2},0)+8(0,\frac{1}{2})+8(0,0), \\
\theta^2,\theta^4 & : & 4(0,0)_{-\frac{1}{3}}+2(1,0)+2(0,1)+
8(\frac{1}{2},\frac{1}{2})+6(\frac{1}{2},0)+6(0,\frac{1}{2})+8(0,0), \\
\theta^3 & : &  8(\frac{1}{2},\frac{1}{2})
+8(\frac{1}{2},0)+8(0,\frac{1}{2})+24(0,0).
\end{array}
\end{equation}
As well as Type IIB string theory, 
states are fewer than the standard. 
In the $\theta^2$+$\theta^4$ sector, 
2(1,0)+2(0,1)+2($\frac{1}{2}$,0)+2(0,$\frac{1}{2}$)+4(0,0)
disappear from the spectrum compared with the $\theta$ sector
of the $T^2$$\times$$T^2/Z_3$ orbifold with $v$=(0,0,$\frac{2}{3}$).  
2(1,0)+2(0,1)+4(0,0) out of these states come from the R-R sectors,
thus we expect the corresponding fractional D-branes \cite{bsf}
also disappear.
In the $\theta^3$ sector, 8($\frac{1}{2}$,0)+8(0,$\frac{1}{2}$)+8(0,0) 
more are necessary to fill 8 vector multiplets.

The R-R states are  bosonic states,
therefore, we expect those also disappear in the spectra
of Type 0 string theories.
In fact, the tacyonic and massless spectrum of Type 0B string theory
on this $T^4/Z_6$ orbifold is   
\begin{equation}
\begin{array}{lcl}
\theta^0 & : & (0,0)_{-\frac{1}{2}}+4(1,0)+4(0,1)+14(0,0),  \\
\theta,\theta^5 & : & 16(0,0)_{-\frac{1}{6}}
+8(1,0)+8(0,1)+32(0,0), \\
\theta^2,\theta^4 & : & 4(0,0)_{-\frac{1}{3}}+4(0,0)_{-\frac{1}{6}}+
8(1,0)+8(0,1)+8(\frac{1}{2},\frac{1}{2})+20(0,0), \\
\theta^3 & : & 8(1,0)+8(0,1)+72(0,0).
\end{array}
\end{equation}
And also, that of Type 0A string theory is 
\begin{equation}
\begin{array}{lcl}
\theta^0 & : & (0,0)_{-\frac{1}{2}}+8(\frac{1}{2},\frac{1}{2})+6(0,0), \\
\theta,\theta^5 & : & 16(0,0)_{-\frac{1}{6}}
+16(\frac{1}{2},\frac{1}{2})+16(0,0), \\
\theta^2,\theta^4 & : & 4(0,0)_{-\frac{1}{3}}+4(0,0)_{-\frac{1}{6}}+
4(1,0)+4(0,1)+16(\frac{1}{2},\frac{1}{2})+12(0,0), \\
\theta^3 & : & 16(\frac{1}{2},\frac{1}{2})+56(0,0).
\end{array}
\end{equation}
We can confirm that 
doubled R-R states as compared with Type IIB/IIA string theory 
disappear in the $\theta^2$+$\theta^4$ sector of Type 0B/0A string theory.
That is, 8($\frac{1}{2}$,$\frac{1}{2}$) disappear in Type 0B string theory,
4(1,0)+4(0,1)+8(0,0) in Type 0A.
The $\theta^3$ sector of both Type 0 string theories
has 8 fewer (0,0) than 
the standard i.e. the $\theta$ sector of the $T^4/Z_2$ orbifold with 
$v$=(0,$\frac{1}{2}$,$-\frac{1}{2}$).


We consider the $T^4/Z_6$ orbifold with 
$v$=(0,$\frac{1}{6}$,$\frac{5}{6}$).  
We find the tachyonic and massless spectrum of Type IIB string theory
on this $T^4/Z_6$ orbifold is
\begin{equation}
\begin{array}{lcl}
\theta^0 & : & 4(1,0)+2(0,1)+14(0,0), \\
\theta,\theta^5 & : & 2(0,0)_{-\frac{1}{3}}+8(0,0)_{-\frac{1}{6}}+
2(1,0)+12(\frac{1}{2},0)+20(0,0), \\
\theta^2,\theta^4 & : & 10(0,1)+32(0,\frac{1}{2})+50(0,0), \\
\theta^3 & : &  6(1,0)+20(\frac{1}{2},0)+28(0,0).
\end{array}
\end{equation}
In spite of 2$v$ and 3$v$ being supersymmetric,
the $\theta^2$+$\theta^4$ and $\theta^3$ sectors do not form
the tensor multiplets.
The $\theta^2$+$\theta^4$ and $\theta^3$ sectors want
8(0,$\frac{1}{2}$) and 4($\frac{1}{2}$,0)+2(0,0) to fill 
the tensor multiplets, respectively.
And also, we find the tachyonic and massless spectrum
of Type IIA string theory is
\begin{equation}
\begin{array}{lcl}
\theta^0 & : & 6(\frac{1}{2},\frac{1}{2})+8(0,0), \\
\theta,\theta^5 & : & 2(0,0)_{-\frac{1}{3}}+8(0,0)_{-\frac{1}{6}}
+2(\frac{1}{2},\frac{1}{2})+
6(\frac{1}{2},0)+6(0,\frac{1}{2})+18(0,0), \\
\theta^2,\theta^4 & : & 10(\frac{1}{2},\frac{1}{2})
+16(\frac{1}{2},0)+16(0,\frac{1}{2})+40(0,0), \\
\theta^3 & : &  6(\frac{1}{2},\frac{1}{2})
+10(\frac{1}{2},0)+10(0,\frac{1}{2})+22(0,0).
\end{array}
\end{equation}
Similar to Type IIB string theory, in order to fill the vector multiplets,
the $\theta^2$+$\theta^4$ sector wants 
4($\frac{1}{2}$,0)+4(0,$\frac{1}{2}$) and the $\theta^3$ sector does
2($\frac{1}{2}$,0)+2(0,$\frac{1}{2}$)+2(0,0).

That the twisted sector with a supersymmetric twist vector 
does not form supersymmetric multiplets
is the same as the $\theta^2$ sector of the $T^4/Z_4$ orbifold 
with $v$=(0,$\frac{1}{4}$,$\frac{3}{4}$).
However, there is a different thing in the $\theta^3$ sector of 
each Type II string theory on this $T^4/Z_6$ orbifold.
That is, 2 fewer (0,0) coming from the \textbf{NS-NS} sector.
Since the NS-NS sector is the bosonic one,
we can anticipate the 2(0,0) also disappear in the spectra of Type 0
string theories.
Then, we shall confirm that as follows.
The tachyonic and massless spectrum of Type 0B string
theory on this $T^4/Z_6$ orbifold is
\begin{equation}
\begin{array}{lcl}
\theta^0 & : & (0,0)_{-\frac{1}{2}}+6(1,0)+6(0,1)+20(0,0),  \\
\theta,\theta^5 & : & 2(0,0)_{-\frac{1}{3}}+8(0,0)_{-\frac{1}{6}}
+2(1,0)+2(0,1)+30(0,0), \\
\theta^2,\theta^4 & : & 10(0,0)_{-\frac{1}{6}}+
10(1,0)+10(0,1)+60(0,0), \\
\theta^3 & : & 6(1,0)+6(0,1)+58(0,0),
\end{array}
\end{equation}
and that of Type 0A string theory is
\begin{equation}
\begin{array}{lcl}
\theta^0 & : & (0,0)_{-\frac{1}{2}}+12(\frac{1}{2},\frac{1}{2})+8(0,0), \\
\theta,\theta^5 & : & 2(0,0)_{-\frac{1}{3}}+8(0,0)_{-\frac{1}{6}}
+4(\frac{1}{2},\frac{1}{2})+26(0,0), \\
\theta^2,\theta^4 & : & 10(0,0)_{-\frac{1}{6}}
+20(\frac{1}{2},\frac{1}{2})+40(0,0), \\
\theta^3 & : & 12(\frac{1}{2},\frac{1}{2})+46(0,0).
\end{array}
\end{equation}
We notice that the $\theta^3$ sector of Type 0B/0A string theory
wants 2(0,0) from the NS-NS sector as we expected.
Therefore, we cannot identify two $T^4/Z_6$ orbifolds
with each other: one with
$v$=(0,$\frac{1}{6}$,$\frac{5}{6}$) and the other 
with $v$=(0,$\frac{1}{6}$,$-\frac{1}{6}$).
This makes a contrast with the relation between two $T^4/Z_4$ orbifolds:
one with $v$=(0,$\frac{1}{4}$,$\frac{3}{4}$) and the other with 
$v$=(0,$\frac{1}{4}$,$-\frac{1}{4}$).


We consider the $T^4/Z_8$ orbifold with $v$=(0,$\frac{1}{8}$,$\frac{3}{8}$). 
The tachyonic and massless spectrum of Type IIB string theory 
on this orbifold is
\begin{equation}
\begin{array}{lcl}
\theta^0 & : & 2(1,0)+2(0,1)+4(0,\frac{1}{2})+8(0,0), \\
\theta,\theta^7 & : & 4(0,0)_{-\frac{1}{8}}+
4(1,0)+8(\frac{1}{2},0)+8(0,0), \\
\theta^2,\theta^6 & : & 6(0,0)_{-\frac{1}{4}}+
6(1,0)+30(0,0), \\
\theta^3,\theta^5 & : & 4(0,0)_{-\frac{1}{4}}+4(0,0)_{-\frac{1}{8}}+ 
4(0,1)+8(0,\frac{1}{2})+8(0,0), \\
\theta^4 & : & 6(0,1)+12(0,\frac{1}{2})+26(0,0).
\end{array}
\end{equation}
By comparison with the standard, i.e. the $\theta$ sector of the
$T^4/Z_4$ orbifold with $v$=(0,$\frac{1}{4}$,$\frac{3}{4}$), 
the $\theta^2$+$\theta^6$ sector lacks 24($\frac{1}{2}$,0).
In addition, we notice that
this sector has only bosonic states 
in spite of Type IIB string theory.
Furthermore the $\theta^4$ sector lacks 12(0,$\frac{1}{2}$)+4(0,0)
to fill the 6 tensor multiplets.
We also find tachyonic and massless spectrum of Type IIA 
string theory is 
\begin{equation}
\begin{array}{lcl}
\theta^0 & : & 4(\frac{1}{2},\frac{1}{2})+2(0,\frac{1}{2})+
2(\frac{1}{2},0)+4(0,0), \\
\theta,\theta^7 & : & 4(0,0)_{-\frac{1}{8}}
+4(\frac{1}{2},\frac{1}{2})+
4(\frac{1}{2},0)+4(0,\frac{1}{2})+4(0,0), \\
\theta^2,\theta^6 & : & 6(0,0)_{-\frac{1}{4}}
+6(\frac{1}{2},\frac{1}{2})+24(0,0), \\
\theta^3,\theta^5 & : &  4(0,0)_{-\frac{1}{4}}+4(0,0)_{-\frac{1}{8}}
+4(\frac{1}{2},\frac{1}{2})
+4(\frac{1}{2},0)+4(0,\frac{1}{2})+4(0,0), \\
\theta^4 & : & 6(\frac{1}{2},\frac{1}{2})+6(\frac{1}{2},0)
+6(0,\frac{1}{2})+20(0,0).
\end{array}
\end{equation}
Similar to Type IIB string theory,
the $\theta^2$+$\theta^6$ sector lacks
12($\frac{1}{2}$,0)+12(0,$\frac{1}{2}$), thus this sector
is bosonic. 
And the $\theta^4$ sector wants 6($\frac{1}{2}$,0)+6(0,$\frac{1}{2}$)+4(0,0)
to fill the 6 vector multiplets.
 
We also consider Type 0 string theories.
The tachyonic and massless spectrum of Type 0B string theory on this
$T^4/Z_8$ orbifold is
\begin{equation}
\begin{array}{lcl}
\theta^0 & : & (0,0)_{-\frac{1}{2}}+4(1,0)+4(0,1)+12(0,0),  \\
\theta,\theta^7 & : & 4(0,0)_{-\frac{1}{4}}+8(0,0)_{-\frac{1}{8}}
+4(1,0)+4(0,1)+16(0,0), \\
\theta^2,\theta^6 & : & 6(0,0)_{-\frac{1}{4}}+
6(1,0)+6(0,1)+60(0,0), \\
\theta^3,\theta^5 & : & 4(0,0)_{-\frac{1}{4}}+8(0,0)_{-\frac{1}{8}}+
4(1,0)+4(0,1)+16(0,0),\\
\theta^4 & : & 6(1,0)+6(0,1)+56(0,0),
\end{array}
\end{equation}
and that of Type 0A string theory is
\begin{equation}
\begin{array}{lcl}
\theta^0 & : & (0,0)_{-\frac{1}{2}}+8(\frac{1}{2},\frac{1}{2})+4(0,0), \\
\theta,\theta^7 & : & 4(0,0)_{-\frac{1}{4}}+8(0,0)_{-\frac{1}{8}}
+8(\frac{1}{2},\frac{1}{2})+8(0,0), \\
\theta^2,\theta^6 & : & 6(0,0)_{-\frac{1}{4}}+
12(\frac{1}{2},\frac{1}{2})+48(0,0), \\
\theta^3,\theta^5 & : & 4(0,0)_{-\frac{1}{4}}+8(0,0)_{-\frac{1}{8}}+
8(\frac{1}{2},\frac{1}{2})+8(0,0),\\
\theta^4 & : & 12(\frac{1}{2},\frac{1}{2})+44(0,0).
\end{array}
\end{equation}
The $\theta^4$ sector of Type 0B/0A string theory has
4 fewer (0,0) than
the standard, i.e. the $\theta$ sector of the $T^4/Z_2$ orbifold with 
$v$=(0,$\frac{1}{2}$,$-\frac{1}{2}$). 


Let us take the non-supersymmetric $T^4/Z_{10}$ orbifold with 
$v$=(0,$\frac{1}{10}$,$\frac{3}{10}$).  
We find the tachyonic and massless spectrum of Type IIB string theory
on this orbifold is
\begin{equation}
\begin{array}{lcl}
\theta^0 & : & 2(1,0)+2(0,1)+4(0,\frac{1}{2})+8(0,0), \\
\theta,\theta^9 & : & 2(0,0)_{-\frac{1}{10}}+
2(1,0)+4(\frac{1}{2},0)+4(0,0), \\
\theta^2,\theta^8 & : & 6(0,0)_{-\frac{1}{5}}+6(1,0)+12(0,0), \\
\theta^3,\theta^7 & : & 2(0,0)_{-\frac{3}{10}}+2(0,0)_{-\frac{1}{5}}+ 
2(0,0)_{-\frac{1}{10}}+2(1,0)+8(\frac{1}{2},0)+10(0,0), \\
\theta^4,\theta^6 & : & 6(0,0)_{-\frac{1}{5}}+
6(0,1)+12(0,\frac{1}{2})+12(0,0),\\
\theta^5 & : & 4(0,1)+12(0,\frac{1}{2})+18(0,0).
\end{array}
\end{equation}
In comparison with the standard, 
i.e. the twisted sectors of the $T^4/Z_5$ orbifold with 
$v$=(0,$\frac{1}{5}$,$\frac{3}{5}$),
the $\theta^4$+$\theta^6$ sector lacks no state,
in contrast, the $\theta^2$+$\theta^8$ sector does
12($\frac{1}{2}$,0); thus this sector is bosonic.
In spite of 5$v$ being supersymmetric,
the $\theta^5$ sector lacks 4(0,$\frac{1}{2}$)+2(0,0)
to fill the 4 tensor multiplets.
And also, we find the tachyonic and massless spectrum of Type IIA 
string theory is  
\begin{equation}
\begin{array}{lcl}
\theta^0 & : & 4(\frac{1}{2},\frac{1}{2})+2(0,\frac{1}{2})+
2(\frac{1}{2},0)+4(0,0), \\
\theta,\theta^9 & : & 2(0,0)_{-\frac{1}{10}}
+2(\frac{1}{2},\frac{1}{2})+
2(\frac{1}{2},0)+2(0,\frac{1}{2})+2(0,0), \\
\theta^2,\theta^8 & : & 6(0,0)_{-\frac{1}{5}}
+6(\frac{1}{2},\frac{1}{2})+6(0,0), \\
\theta^3,\theta^7 & : & 2(0,0)_{-\frac{3}{10}}+2(0,0)_{-\frac{1}{5}} 
+2(0,0)_{-\frac{1}{10}}
+2(\frac{1}{2},\frac{1}{2})
+4(\frac{1}{2},0)+4(0,\frac{1}{2})+8(0,0), \\
\theta^4,\theta^6 & : & 6(0,0)_{-\frac{1}{5}}+
6(\frac{1}{2},\frac{1}{2})+6(\frac{1}{2},0)+6(0,\frac{1}{2})+6(0,0),\\
\theta^5 & : & 4(\frac{1}{2},\frac{1}{2})
+6(\frac{1}{2},0)+6(0,\frac{1}{2})+14(0,0).
\end{array}
\end{equation}
Parallel to Type IIB string theory,
the $\theta^2$+$\theta^8$ sector, which lacks 
6($\frac{1}{2}$,0)+6(0,$\frac{1}{2}$), is bosonic and
the $\theta^5$ sector lacks 
2($\frac{1}{2}$,0)+2(0,$\frac{1}{2}$)+2(0,0) to fill 
the 4 vector multiplets.

In addition, we find the tachyonic and massless spectrum 
of Type 0B string theory on this $T^4/Z_{10}$ orbifold is
\begin{equation}
\begin{array}{lcl}
\theta^0 & : & (0,0)_{-\frac{1}{2}}+4(1,0)+4(0,1)+12(0,0),  \\
\theta,\theta^9 & : & 2(0,0)_{-\frac{3}{10}}+2(0,0)_{-\frac{1}{5}}+ 
4(0,0)_{-\frac{1}{10}}+2(1,0)+2(0,1)+14(0,0), \\
\theta^2,\theta^8 & : & 6(0,0)_{-\frac{1}{5}}+6(0,0)_{-\frac{1}{10}}+
6(1,0)+6(0,1)+18(0,0), \\
\theta^3,\theta^7 & : & 2(0,0)_{-\frac{3}{10}}+2(0,0)_{-\frac{1}{5}}+
4(0,0)_{-\frac{1}{10}}+2(1,0)+2(0,1)+14(0,0),\\
\theta^4,\theta^6 & : &  6(0,0)_{-\frac{1}{5}}+6(0,0)_{-\frac{1}{10}}+
6(1,0)+6(0,1)+18(0,0), \\
\theta^5 & : & 4(1,0)+4(0,1)+38(0,0), 
\end{array}
\end{equation}
and that of Type 0A string theory is
\begin{equation}
\begin{array}{lcl}
\theta^0 & : & (0,0)_{-\frac{1}{2}}+8(\frac{1}{2},\frac{1}{2})+4(0,0), \\
\theta,\theta^9 & : & 2(0,0)_{-\frac{3}{10}}+2(0,0)_{-\frac{1}{5}}
+4(0,0)_{-\frac{1}{10}}+4(\frac{1}{2},\frac{1}{2})+10(0,0), \\
\theta^2,\theta^8 & : & 6(0,0)_{-\frac{1}{5}}+6(0,0)_{-\frac{1}{10}}
+12(\frac{1}{2},\frac{1}{2})+6(0,0), \\
\theta^3,\theta^7 & : & 2(0,0)_{-\frac{3}{10}}+2(0,0)_{-\frac{1}{5}}+
4(0,0)_{-\frac{1}{10}}+4(\frac{1}{2},\frac{1}{2})+10(0,0),\\
\theta^4,\theta^6  & : & 6(0,0)_{-\frac{1}{5}}+6(0,0)_{-\frac{1}{10}}+
12(\frac{1}{2},\frac{1}{2})+6(0,0),\\
\theta^5 & : & 8(\frac{1}{2},\frac{1}{2})+30(0,0).
\end{array}
\end{equation}
Being contrastive to Type II string theories on this orbifold,
in both Type 0 string theories,
there is no shortage of the states in the twisted sectors
except for the $\theta^5$ sector, which has 2 fewer (0,0) than 
the standard, i.e. the $\theta$ sector of the $T^4/Z_2$ orbifold
with $v$=(0,$\frac{1}{2}$,$-\frac{1}{2}$).

The two following $T^4/Z_{12}$ orbifolds have the same orders, but a slight
difference appear between each spectrum of a string theory on two orbifolds. 
First, we consider the non-supersymmetric $T^4/Z_{12}$ orbifold 
with $v$=(0,$\frac{1}{12}$,$\frac{5}{12}$).  
We find the tachyonic and massless spectrum of Type IIB string theory 
on this $T^4/Z_{12}$ orbifold is
\begin{equation}
\begin{array}{lcl}
\theta^0 & : & 2(1,0)+2(0,1)+8(0,0), \\
\theta,\theta^5,\theta^7,\theta^{11} & : & 
4(0,0)_{-\frac{1}{6}}+4(0,0)_{-\frac{1}{12}}
+4(1,0)+8(\frac{1}{2},0)+8(0,0), \\
\theta^2,\theta^{10} & : & 2(0,0)_{-\frac{1}{3}}+
8(0,0)_{-\frac{1}{6}}+2(1,0)+20(0,0), \\
\theta^3,\theta^9 & : & 4(0,0)_{-\frac{1}{4}}+4(0,1)+20(0,0), \\
\theta^4,\theta^8 & : & 6(0,1)+16(0,\frac{1}{2})+26(0,0),\\
\theta^6 & : & 4(1,0)+8(\frac{1}{2},0)+18(0,0).
\end{array}
\end{equation}
We notice at once that the $\theta^2$+$\theta^{10}$ and
the $\theta^3$+$\theta^9$ sectors are bosonic.
The former wants 12($\frac{1}{2}$,0) 
to fill the standard number ratio of the states, i.e.
the ratio in the $\theta$ sector 
of the $T^4/Z_{6}$ orbifold with  
$v$=(0,$\frac{1}{6}$,$\frac{5}{6}$).
And the latter also wants 16(0,$\frac{1}{2}$)
to fill the standard, i.e.
that in the $\theta$ sector of the $T^4/Z_{4}$ orbifold with
$v$=(0,$\frac{1}{4}$,$-\frac{3}{4}$).
Though 4$v$ and 6$v$ are supersymmetric twist vectors, 
the $\theta^4$+$\theta^8$ sector lacks 
8(0,$\frac{1}{2}$)+4(0,0) and the $\theta^6$ sector does
8($\frac{1}{2}$,0)+2(0,0) to fill the tensor multiplets.
The similar things also appear in the tachyonic and massless spectrum 
of Type IIA string theory as follows:
\begin{equation}
\begin{array}{lcl}
\theta^0 & : & 4(\frac{1}{2},\frac{1}{2})+4(0,0), \\
\theta,\theta^5,\theta^7,\theta^{11} & : & 4(0,0)_{-\frac{1}{6}}
+4(0,0)_{-\frac{1}{12}}+4(\frac{1}{2},\frac{1}{2})+
4(\frac{1}{2},0)+4(0,\frac{1}{2})+4(0,0), \\
\theta^2,\theta^{10} & : & 2(0,0)_{-\frac{1}{3}}+8(0,0)_{-\frac{1}{6}}
+2(\frac{1}{2},\frac{1}{2})+18(0,0), \\
\theta^3,\theta^9 & : & 4(0,0)_{-\frac{1}{4}}
+4(\frac{1}{2},\frac{1}{2})+16(0,0), \\
\theta^4,\theta^8 & : & 6(\frac{1}{2},\frac{1}{2})
+8(\frac{1}{2},0)+8(0,\frac{1}{2})+20(0,0),\\
\theta^6 & : & 4(\frac{1}{2},\frac{1}{2})
+4(\frac{1}{2},0)+4(0,\frac{1}{2})+14(0,0).
\end{array}
\end{equation}
As well as Type IIB string theory,
the $\theta^2$+$\theta^{10}$ and the $\theta^3$+$\theta^9$ sector 
are bosonic.
By comparison with the standard,
there disappear 6($\frac{1}{2}$,0)+6(0,$\frac{1}{2}$) in the
former and 8($\frac{1}{2}$,0)+8(0,$\frac{1}{2}$) in the latter.
Added to this, the $\theta^4$+$\theta^8$ sector wants 
4($\frac{1}{2}$,0)+4(0,$\frac{1}{2}$)+4(0,0) and 
the $\theta^6$ sector dose 4($\frac{1}{2}$,0)+4(0,$\frac{1}{2}$)+2(0,0)
to fill the vector multiplets.

In addition, we consider Type 0 string theories on this $T^4/Z_{12}$
orbifold.
The tachyonic and massless spectrum of Type 0B string theory on this 
orbifold is
\begin{equation}
\begin{array}{lcl}
\theta^0 & : & (0,0)_{-\frac{1}{2}}+4(1,0)+4(0,1)+12(0,0),  \\
\theta,\theta^5,\theta^7,\theta^{11} & : & 
4(0,0)_{-\frac{1}{4}}+8(0,0)_{-\frac{1}{6}}+8(0,0)_{-\frac{1}{12}}
+4(1,0)+4(0,1)+16(0,0), \\
\theta^2,\theta^{10} & : & 2(0,0)_{-\frac{1}{3}}+8(0,0)_{-\frac{1}{6}}+
2(1,0)+2(0,1)+30(0,0), \\
\theta^3,\theta^9 & : & 4(0,0)_{-\frac{1}{4}}+4(1,0)+4(0,1)+40(0,0),\\
\theta^4,\theta^8 & : &  6(0,0)_{-\frac{1}{6}}+6(1,0)+6(0,1)+32(0,0), \\
\theta^6 & : & 4(1,0)+4(0,1)+38(0,0), 
\end{array}
\end{equation}
and that of Type 0A string theory is
\begin{equation}
\begin{array}{lcl}
\theta^0 & : & (0,0)_{-\frac{1}{2}}+8(\frac{1}{2},\frac{1}{2})+4(0,0), \\
\theta,\theta^5,\theta^7,\theta^{11} & : & 
4(0,0)_{-\frac{1}{4}}+8(0,0)_{-\frac{1}{6}}
+8(0,0)_{-\frac{1}{12}}+8(\frac{1}{2},\frac{1}{2})+8(0,0), \\
\theta^2,\theta^{10} & : & 2(0,0)_{-\frac{1}{3}}+8(0,0)_{-\frac{1}{6}}+
4(\frac{1}{2},\frac{1}{2})+26(0,0), \\
\theta^3,\theta^9 & : & 4(0,0)_{-\frac{1}{4}}+
8(\frac{1}{2},\frac{1}{2})+32(0,0),\\
\theta^4,\theta^8 & : & 6(0,0)_{-\frac{1}{6}}+
12(\frac{1}{2},\frac{1}{2})+20(0,0),\\
\theta^6 & : & 8(\frac{1}{2},\frac{1}{2})+30(0,0).
\end{array}
\end{equation}
Several states, corresponding to disappearing bosonic states
in Type II string theories, disappear in both Type 0 string theories. 
The  $\theta^4$+$\theta^8$ sector wants 4(0,0) 
to fill the standard number ratio of the states, 
i.e. the ratio in the $\theta$ sector of the $T^3/Z_3$ orbifold with 
$v$=(0,$\frac{1}{3}$,$-\frac{1}{3}$).   
And also, the  $\theta^6$ sector wants 2(0,0) 
to fill the standard number ratio of the states,
i.e. that in the $\theta$ sector of the $T^4/Z_2$ orbifold with 
$v$=(0,$\frac{1}{2}$,$-\frac{1}{2}$).


Second, and finally in this subsection,
we consider the non-supersymmetric $T^4/Z_{12}$ orbifold 
with $v$=(0,$\frac{1}{12}$,$\frac{7}{12}$).
The tachyonic and massless spectrum of Type IIB string theory on
this orbifold is
\begin{equation}
\begin{array}{lcl}
\theta^0 & : & 2(1,0)+2(0,1)+8(0,0), \\
\theta,\theta^5,\theta^7,\theta^{11} & : & 
4(0,0)_{-\frac{1}{4}}+4(0,0)_{-\frac{1}{6}}+4(0,0)_{-\frac{1}{12}}
+4(1,0)+8(\frac{1}{2},0)+8(0,0), \\
\theta^2,\theta^{10} & : & 2(0,0)_{-\frac{1}{3}}+
8(0,0)_{-\frac{1}{6}}+2(0,1)+20(0,0), \\
\theta^3,\theta^9 & : & 4(0,1)+8(0,\frac{1}{2})+16(0,0), \\
\theta^4,\theta^8 & : & 6(1,0)+16(\frac{1}{2},0)+26(0,0),\\
\theta^6 & : & 4(0,1)+12(0,\frac{1}{2})+18(0,0).
\end{array}
\end{equation}
We notice only the $\theta^2$+$\theta^{10}$ sector is bosonic,
which lacks 12(0,$\frac{1}{2}$) as compared with the $\theta$ sector 
of the $T^4/Z_6$ orbifold with 
$v$=(0,$\frac{1}{6}$,$-\frac{5}{6}$), in contrast,
Type II string theory on the previous orbifold  
has more bosonic sectors.
3$v$, 4$v$ and 6$v$ are supersymmetric twist vectors,
but the $\theta^3$+$\theta^9$, $\theta^4$+$\theta^8$ and $\theta^6$
sectors want 8(0,$\frac{1}{2}$)+4(0,0), 8($\frac{1}{2}$,0)+4(0,0)
and 4(0,$\frac{1}{2}$)+2(0,0) respectively to fill the tensor 
multiplets.
Similar things also appear in the tachyonic and massless 
spectrum of Type IIA string theory as follows:
\begin{equation}
\begin{array}{lcl}
\theta^0 & : & 4(\frac{1}{2},\frac{1}{2})+4(0,0), \\
\theta,\theta^5,\theta^7,\theta^{11} & : & 4(0,0)_{-\frac{1}{4}}
+4(0,0)_{-\frac{1}{6}}+4(0,0)_{-\frac{1}{12}}
+4(\frac{1}{2},\frac{1}{2})+
4(\frac{1}{2},0)+4(0,\frac{1}{2})+4(0,0), \\
\theta^2,\theta^{10} & : & 2(0,0)_{-\frac{1}{3}}+8(0,0)_{-\frac{1}{6}}
+2(\frac{1}{2},\frac{1}{2})+18(0,0), \\
\theta^3,\theta^9 & : & 4(\frac{1}{2},\frac{1}{2})+
4(\frac{1}{2},0)+4(0,\frac{1}{2})+12(0,0), \\
\theta^4,\theta^8 & : & 6(\frac{1}{2},\frac{1}{2})
+8(\frac{1}{2},0)+8(0,\frac{1}{2})+20(0,0),\\
\theta^6 & : & 4(\frac{1}{2},\frac{1}{2})
+6(\frac{1}{2},0)+6(0,\frac{1}{2})+14(0,0).
\end{array}
\end{equation}
The $\theta^2$+$\theta^{10}$ sector lacks 
6($\frac{1}{2}$,0)+6(0,$\frac{1}{2}$), therefore, it is bosonic sector.
And also, the $\theta^3$+$\theta^9$, $\theta^4$+$\theta^8$ and $\theta^6$
sectors want 4($\frac{1}{2}$,0)+4(0,$\frac{1}{2}$)+4(0,0),
4($\frac{1}{2}$,0)+4(0,$\frac{1}{2}$)+4(0,0) 
and 2($\frac{1}{2}$,0)+2(0,$\frac{1}{2}$) +2(0,0)
respectively to fill the vector multiplets.

In addition, we consider Type 0 string theories on this $T^4/Z_{12}$ orbifold.
The tachyonic and massless spectrum of Type 0B string theory on this
orbifold is
\begin{equation}
\begin{array}{lcl}
\theta^0 & : & (0,0)_{-\frac{1}{2}}+4(1,0)+4(0,1)+12(0,0),  \\
\theta,\theta^5,\theta^7,\theta^{11} & : & 
4(0,0)_{-\frac{1}{4}}+8(0,0)_{-\frac{1}{6}}+8(0,0)_{-\frac{1}{12}}
+4(1,0)+4(0,1)+16(0,0), \\
\theta^2,\theta^{10} & : & 2(0,0)_{-\frac{1}{3}}+8(0,0)_{-\frac{1}{6}}+
2(1,0)+2(0,1)+30(0,0), \\
\theta^3,\theta^9 & : & 4(0,0)_{-\frac{1}{4}}+4(1,0)+4(0,1)+36(0,0),\\
\theta^4,\theta^8 & : &  6(0,0)_{-\frac{1}{6}}+6(1,0)+6(0,1)+32(0,0), \\
\theta^6 & : & 4(1,0)+4(0,1)+38(0,0). 
\end{array}
\end{equation}
And that of Type 0A string theory is
\begin{equation}
\begin{array}{lcl}
\theta^0 & : & (0,0)_{-\frac{1}{2}}+8(\frac{1}{2},\frac{1}{2})+4(0,0), \\
\theta,\theta^5,\theta^7,\theta^{11} & : & 
4(0,0)_{-\frac{1}{4}}+8(0,0)_{-\frac{1}{6}}
+8(0,0)_{-\frac{1}{12}}+8(\frac{1}{2},\frac{1}{2})+8(0,0), \\
\theta^2,\theta^{10} & : & 2(0,0)_{-\frac{1}{3}}+8(0,0)_{-\frac{1}{6}}
+4(\frac{1}{2},\frac{1}{2})+26(0,0), \\
\theta^3,\theta^9 & : & 4(0,0)_{-\frac{1}{4}}+
8(\frac{1}{2},\frac{1}{2})+28(0,0),\\
\theta^4,\theta^8 & : & 6(0,0)_{-\frac{1}{6}}+
12(\frac{1}{2},\frac{1}{2})+20(0,0),\\
\theta^6 & : & 8(\frac{1}{2},\frac{1}{2})+30(0,0).
\end{array}
\end{equation}
There appear the same spectrum of each Type 0 string theory
on the $T^4/Z_{12}$ orbifold with 
$v$=(0,$\frac{1}{12}$,$\frac{5}{12}$) except for 
the $\theta^3$+$\theta^9$ sector, which has 4 fewer (0,0)
than the standard, i.e. the number ratio of the states 
in the $\theta$ sector of 
the $T^4/Z_{4}$ orbifold with $v$=(0,$\frac{1}{4}$,$-\frac{1}{4}$).
This difference depends on whether 3$v$ is supersymmetric or not.
Therefore, in Type 0 string theories,
we cannot identify two $T^4/Z_{12}$ orbifolds: 
one with $v$=(0,$\frac{1}{12}$,$\frac{5}{12}$) and the other with
$v$=(0,$\frac{1}{12}$,$\frac{7}{12}$).
This is similar to the relation between two $T^4/Z_6$ orbifolds:
one with $v$=(0,$\frac{1}{6}$,$\frac{5}{6}$) 
and the other with $v$=(0,$\frac{1}{6}$,$-\frac{1}{6}$).
And also, this makes a contrast with the relation between 
two $T^4/Z_4$ orbifolds:
one with $v$=(0,$\frac{1}{4}$,$\frac{3}{4}$) and the other with
$v$=(0,$\frac{1}{4}$,$-\frac{1}{4}$).

\subsection{Orbifolds including the $(-1)^{F_S}$ twist}
\label{t4.2}

In this subsection, we focus on $T^4/Z_N$ orbifolds
including the $(-1)^{F_S}$ twist, therefore,  
we consider only Type 0 string theories on
a orbifold whose twist vector is listed in Table \ref{tab3}.

\begin{table}[htb]
\begin{center}
 \begin{tabular}{|c|c||c|c||c|c|} 
 \hline
  $(v_2,v_3,v_4)$ & $Z_N$(for Type II) & $Z_N$(for Type 0) \\
 \hline
 (0,$\frac{1}{3}$,$\frac{2}{3}$) & $Z_6$ & $Z_3$ \\
 \hline
 (0,$\frac{1}{4}$,$\frac{2}{4}$)=(0,$\frac{1}{4}$,$\frac{1}{2}$) 
& $Z_8$ & $Z_4$ \\  
 \hline
 (0,$\frac{1}{5}$,$\frac{2}{5}$) & $Z_{10}$ & $Z_5$ \\
 \hline 
 (0,$\frac{1}{6}$,$\frac{2}{6}$)=(0,$\frac{1}{6}$,$\frac{1}{3}$) 
& $Z_{12}$ & $Z_{6}$ \\
 (0,$\frac{2}{6}$,$\frac{3}{6}$)=(0,$\frac{1}{3}$,$\frac{1}{2}$) 
& $Z_{12}$ & $Z_{6}$ \\
 \hline 
 (0,$\frac{2}{12}$,$\frac{3}{12}$)=(0,$\frac{1}{6}$,$\frac{1}{4}$) 
& $Z_{24}$ & $Z_{12}$ \\
 (0,$\frac{3}{12}$,$\frac{4}{12}$)=(0,$\frac{1}{4}$,$\frac{1}{3}$) 
& $Z_{24}$ & $Z_{12}$ \\
 \hline
\end{tabular}
\end{center}
\caption{Twist vectors  
including the $(-1)^{F_S}$ twist with two non-zero components.
We take absolute value of two components of the twist vectors.}
\label{tab3}
\end{table}

Some twisted sectors of Type 0 string theories
on this type $T^4/Z_N$ orbifold 
want several states to fill the standard number ratio of the states. 
Moreover all disappearing states come from only
the twisted R-R sector whose twist vector has one non-zero component.
From the orbifolds considered in the previous subsection,
we can give only example in which the R-R states disappear,
i.e. the $\theta^2$+$\theta^4$ sector
of the $T^4/Z_6$ orbifold with 
$v$=(0,$\frac{1}{6}$,$\frac{3}{6}$)=(0,$\frac{1}{6}$,$\frac{1}{2}$).   
In this subsection,
let us describe the spectrum in the same manner as 
the previous subsection, but 
the order of a orbifold indicates that for
Type 0 string theories.

As the first example, let us take
the $T^4/Z_3$ orbifold with $v$=(0,$\frac{1}{3}$,$\frac{2}{3}$).  
The tachyonic and massless spectrum of 
Type 0B string theory on this orbifold is
\begin{equation}
\begin{array}{lcl}
\theta^0 & : & (0,0)_{-\frac{1}{2}}+6(1,0)+6(0,1)+20(0,0),  \\
\theta,\theta^2 & : & 
18(0,0)_{-\frac{1}{6}}+18(1,0)+18(0,1)+108(0,0).
\end{array}
\end{equation}
As indicated in \cite{hepth0202057},
this spectrum is equivalent to that on the $T^4/Z_3$ orbifold with 
$v$=(0,$\frac{1}{3}$,$-\frac{1}{3}$).  
The same thing also appear in that of Type 0A string theory as follows:
\begin{equation}
\begin{array}{lcl}
\theta^0 & : & (0,0)_{-\frac{1}{2}}+12(\frac{1}{2},\frac{1}{2})+8(0,0), \\
\theta,\theta^2 & : & 
18(0,0)_{-\frac{1}{6}}+36(\frac{1}{2},\frac{1}{2})+72(0,0).
\end{array}
\end{equation}
Therefore, for Type 0 string theories
we can identify two $T^4/Z_3$ orbifolds:
one with $v$=(0,$\frac{1}{3}$,$-\frac{1}{3}$) and 
the other with $v$=(0,$\frac{1}{3}$,$\frac{2}{3}$).


Let us consider the $T^4/Z_4$ orbifold with  
$v$=(0,$\frac{1}{4}$,$\frac{2}{4}$)=(0,$\frac{1}{4}$,$\frac{1}{2}$). 
This orbifold has two different twists.
In fact, we find the tachyonic and massless spectrum of Type 0B string 
theory on this orbifold is
\begin{equation}
\begin{array}{lcl}
\theta^0 & : & (0,0)_{-\frac{1}{2}}+4(1,0)+4(0,1)+14(0,0),  \\
\theta,\theta^3 & : & 
32(0,0)_{-\frac{1}{8}}+16(1,0)+16(0,1)+32(0,0), \\
\theta^2 & : & 6(0,0)_{-\frac{1}{4}}+6(1,0)+6(0,1)
+4(\frac{1}{2},\frac{1}{2})+12(0,0).
\end{array}
\end{equation}
Since 2$v$ is (0,$\frac{1}{2}$,0), the $\theta^2$ sector
lacks 8($\frac{1}{2}$,$\frac{1}{2}$) in 
comparison with the standard number ratio of the states, 
i.e. the ratio in the $\theta$ sector of
the $T^2$$\times$$T^2/Z_2$ orbifold with $v$=(0,0,$\frac{1}{2}$).  
As mentioned above, the disappearing states accompany 
the twist vector which has one non-zero component. 
In addition, the disappearing states are
always some ($\frac{1}{2}$,$\frac{1}{2}$) in the twisted sector of 
Type 0B string theory.
Similar thing also appear in the tachyonic and massless spectrum
of Type 0A string theory as follows: 
\begin{equation}
\begin{array}{lcl}
\theta^0 & : & (0,0)_{-\frac{1}{2}}+8(\frac{1}{2},\frac{1}{2})+6(0,0), \\
\theta,\theta^3 & : & 
32(0,0)_{-\frac{1}{8}}+32(\frac{1}{2},\frac{1}{2}), \\
\theta^2 & : &
6(0,0)_{-\frac{1}{4}}+2(1,0)+2(0,1)+12(\frac{1}{2},\frac{1}{2})+4(0,0). 
\end{array}
\end{equation}
Similar to Type 0B,
the $\theta^2$ sector wants 4(1,0)+4(0,1)+8(0,0) in comparison
with the standard. 
Also in the other Type 0A orbifolds including the $(-1)^{F_S}$ twist,
the disappearing states are several times 
\begin{equation}
(1,0)+(0,1)+2(0,0),  
\end{equation}
therefore we shall call this unit \textit{disappearing unit}[DU]. 
That is, 4 DU disappear in the $\theta^2$ sector of this Type 0A orbifold.
Of course, some DU accompany the twist vector with one non-zero component.


Next, we consider 
the $T^4/Z_5$ orbifold with $v$=(0,$\frac{1}{5}$,$\frac{2}{5}$). 
This orbifold includes no $T^2$$\times$$T^2/Z_N$ type twist.
In fact,
the tachyonic and massless spectrum of Type 0B string theory on this
orbifold is
\begin{equation}
\begin{array}{lcl}
\theta^0 & : & (0,0)_{-\frac{1}{2}}+4(1,0)+4(0,1)+12(0,0),  \\
\theta,\theta^2,\theta^3,\theta^4 & : & 
20(0,0)_{-\frac{1}{5}}+20(0,0)_{-\frac{1}{10}}+20(1,0)+20(0,1)+60(0,0).
\end{array}
\end{equation} 
And also, that of Type 0A string theory is 
\begin{equation}
\begin{array}{lcl}
\theta^0 & : & (0,0)_{-\frac{1}{2}}+8(\frac{1}{2},\frac{1}{2})+4(0,0), \\
\theta,\theta^2,\theta^3,\theta^4 & : & 
20(0,0)_{-\frac{1}{5}}+20(0,0)_{-\frac{1}{10}}
+40(\frac{1}{2},\frac{1}{2})+20(0,0).
\end{array}
\end{equation}
Obviously, we obtain the same spectrum of Type 0B/0A 
string theory on the $T^4/Z_5$ orbifold
with $v$=(0,$\frac{1}{5}$,$\frac{3}{5}$):
thus, we can identify $v$=(0,$\frac{1}{5}$,$\frac{2}{5}$) with
$v$=(0,$\frac{1}{5}$,$\frac{3}{5}$) in Type 0 string theories.  


The $T^4/Z_6$ orbifold with
$v$=(0,$\frac{1}{6}$,$\frac{2}{6}$)=(0,$\frac{1}{6}$,$\frac{1}{3}$) 
include three different twists.
The tachyonic and massless spectrum of Type 0B string theory on 
this orbifold is
\begin{equation}
\begin{array}{lcl}
\theta^0 & : & (0,0)_{-\frac{1}{2}}+4(1,0)+4(0,1)+12(0,0),  \\
\theta,\theta^5 & : & 
6(0,0)_{-\frac{1}{4}}+12(0,0)_{-\frac{1}{12}}
+6(1,0)+6(0,1)+12(0,0), \\
\theta^2,\theta^4 & : & 
12(0,0)_{-\frac{1}{6}}+12(1,0)+12(0,1)+72(0,0), \\
\theta^3 & : & 4(0,0)_{-\frac{1}{4}}+
4(1,0)+4(0,1)+4(\frac{1}{2},\frac{1}{2})+8(0,0). 
\end{array}
\end{equation}
Since 3$v$=(0,$\frac{1}{2}$,0),
the $\theta^3$ sector lacks 4($\frac{1}{2}$,$\frac{1}{2}$)
as compared with standard ratio, 
i.e. the number ratio of the state in the $\theta$ sector of 
the $T^2$$\times$$T^2/Z_2$ orbifold with $v$=(0,0,$\frac{1}{2}$).
And also, we find the tachyonic and massless spectrum of Type 0A
string theory is
\begin{equation}
\begin{array}{lcl}
\theta^0 & : & (0,0)_{-\frac{1}{2}}+8(\frac{1}{2},\frac{1}{2})+4(0,0), \\
\theta,\theta^5 & : & 
6(0,0)_{-\frac{1}{4}}+12(0,0)_{-\frac{1}{12}}
+12(\frac{1}{2},\frac{1}{2}), \\
\theta^2,\theta^4 & : & 
12(0,0)_{-\frac{1}{6}}+24(\frac{1}{2},\frac{1}{2})+48(0,0), \\
\theta^3 & : & 
4(0,0)_{-\frac{1}{4}}+
2(1,0)+2(0,1)+8(\frac{1}{2},\frac{1}{2})+4(0,0).
\end{array}
\end{equation}
Similar to Type 0B string theory, the $\theta^3$ sector lacks 2 DU
as compared with standard.


The orbifold with 
$v$=(0,$\frac{2}{6}$,$\frac{3}{6}$)=(0,$\frac{1}{3}$,$\frac{1}{2}$)
is also a $T^4/Z_6$ orbifold.
The tachyonic and massless spectrum of Type 0B string theory 
on this orbifold is
\begin{equation}
\begin{array}{lcl}
\theta^0 & : & (0,0)_{-\frac{1}{2}}+4(1,0)+4(0,1)+14(0,0),  \\
\theta,\theta^5 & : & 
48(0,0)_{-\frac{1}{12}}+24(1,0)+24(0,1)+48(0,0), \\
\theta^2,\theta^4 & : & 
6(0,0)_{-\frac{1}{3}}+6(0,0)_{-\frac{1}{6}}+12(1,0)+12(0,1)+30(0,0), \\
\theta^3 & : & 8(0,0)_{-\frac{1}{4}}+8(1,0)+8(0,1)+16(0,0). 
\end{array}
\end{equation}
2$v$ and 3$v$ are one non-zero component vectors,
therefore the $\theta^2$+$\theta^4$ and $\theta^3$ sectors want some
($\frac{1}{2}$,$\frac{1}{2}$).
The former wants 24($\frac{1}{2}$,$\frac{1}{2}$)
as compared with the $\theta$ sector of 
the $T^2$$\times$$T^2/Z_3$ orbifold with $v$=(0,0,$\frac{1}{3}$), 
and the latter does 
16($\frac{1}{2}$,$\frac{1}{2}$) as compared with 
the $\theta$ sector of the $T^2$$\times$$T^2/Z_2$ 
orbifold with $v$=(0,0,$\frac{1}{2}$). 
We also find the tachyonic and massless spectrum of Type 0A string theory
is
\begin{equation}
\begin{array}{lcl}
\theta^0 & : & (0,0)_{-\frac{1}{2}}+8(\frac{1}{2},\frac{1}{2})+6(0,0), \\
\theta,\theta^5 & : & 48(0,0)_{-\frac{1}{12}}
+48(\frac{1}{2},\frac{1}{2}), \\
\theta^2,\theta^4 & : & 
6(0,0)_{-\frac{1}{3}}+6(0,0)_{-\frac{1}{6}}
+24(\frac{1}{2},\frac{1}{2})+6(0,0), \\
\theta^3 & : & 
8(0,0)_{-\frac{1}{4}}+16(\frac{1}{2},\frac{1}{2}).
\end{array}
\end{equation}
As well as Type 0B string theory,
the $\theta^2$+$\theta^4$ and $\theta^3$ sectors
want 12 DU and 8 DU respectively.

The last two orbifolds are $T^4/Z_{12}$ orbifolds.
Each of these has five different twists.
First, we consider the $T^4/Z_{12}$ orbifold with 
$v$=(0,$\frac{2}{12}$,$\frac{3}{12}$)=(0,$\frac{1}{6}$,$\frac{1}{4}$). 
The tachyonic and massless spectrum of Type 0B string theory 
on this orbifold is
\begin{equation}
\begin{array}{lcl}
\theta^0 & : & (0,0)_{-\frac{1}{2}}+4(1,0)+4(0,1)+12(0,0),  \\
\theta,\theta^5,\theta^7,\theta^{11} & : & 
8(0,0)_{-\frac{7}{24}}+8(0,0)_{-\frac{1}{8}}+16(0,0)_{-\frac{1}{24}}
+8(1,0)+8(0,1)+16(0,0), \\
\theta^2,\theta^{10} & : & 
28(0,0)_{-\frac{1}{12}}+14(1,0)+14(0,1)+28(0,0), \\
\theta^3,\theta^9 & : & 16(0,0)_{-\frac{1}{4}}+8(1,0)+8(0,1)+16(0,0), \\
\theta^4,\theta^8 & : & 4(0,0)_{-\frac{1}{3}}+4(0,0)_{-\frac{1}{6}}+
8(1,0)+8(0,1)+20(0,0), \\
\theta^6 & : & 6(0,0)_{-\frac{1}{4}}+6(1,0)+6(0,1)+12(0,0). 
\end{array}
\end{equation}
4$v$ and 6$v$ are one non-zero component vectors,
therefore there disappear some ($\frac{1}{2}$,$\frac{1}{2}$).
The $\theta^4$+$\theta^8$ sector has 16 fewer 
($\frac{1}{2}$,$\frac{1}{2}$) than the standard number ratio of
the states, i.e. the ratio in the $\theta$ sector of 
the $T^2$$\times$$T^2/Z_3$ orbifold with $v$=(0,0,$\frac{1}{3}$).
And the $\theta^3$+$\theta^9$ sector has 12 fewer 
($\frac{1}{2}$,$\frac{1}{2}$) than the standard ratio, i.e. 
the ratio in the $\theta$ sector of the $T^2$$\times$$T^2/Z_2$
orbifold with $v$=(0,0,$\frac{1}{2}$).
We also find the tachyonic and massless spectrum of 
Type 0A string theory is
\begin{equation}
\begin{array}{lcl}
\theta^0 & : & (0,0)_{-\frac{1}{2}}+8(\frac{1}{2},\frac{1}{2})+4(0,0), \\
\theta,\theta^5,\theta^7,\theta^{11} & : & 8(0,0)_{-\frac{7}{24}}
+8(0,0)_{-\frac{1}{8}}+16(0,0)_{-\frac{1}{24}}+16(\frac{1}{2},\frac{1}{2}),\\
\theta^2,\theta^{10} & : & 
28(0,0)_{-\frac{1}{12}}+28(\frac{1}{2},\frac{1}{2}), \\
\theta^3,\theta^9 & : & 
16(0,0)_{-\frac{1}{4}}+16(\frac{1}{2},\frac{1}{2}), \\
\theta^4,\theta^8 & : &  
4(0,0)_{-\frac{1}{3}}+4(0,0)_{-\frac{1}{6}}
+16(\frac{1}{2},\frac{1}{2})+4(0,0), \\
\theta^6 & : & 6(0,0)_{-\frac{1}{4}}+12(\frac{1}{2},\frac{1}{2}). 
\end{array}
\end{equation}
Similar to Type 0B string theory,
the $\theta^4$+$\theta^8$ sector lacks 8 DU and
the $\theta^3$+$\theta^9$ sector does 6 DU 
as compared with the standard number ratio of the states. 


Second, we consider the $T^4/Z_{12}$ orbifold with 
$v$=(0,$\frac{3}{12}$,$\frac{4}{12}$)=(0,$\frac{1}{4}$,$\frac{1}{3}$).
The tachyonic and massless spectrum of Type 0B string theory on 
this orbifold is
\begin{equation}
\begin{array}{lcl}
\theta^0 & : & (0,0)_{-\frac{1}{2}}+4(1,0)+4(0,1)+12(0,0),  \\
\theta,\theta^5,\theta^7,\theta^{11} & : & 
24(0,0)_{-\frac{5}{24}}+24(0,0)_{-\frac{1}{24}}
+24(1,0)+24(0,1)+48(0,0), \\
\theta^2,\theta^{10} & : & 
36(0,0)_{-\frac{1}{12}}+18(1,0)+18(0,1)+36(0,0), \\
\theta^3,\theta^9 & : & 4(0,0)_{-\frac{3}{8}}+8(0,0)_{-\frac{1}{8}}
+8(1,0)+8(0,1)+16(0,0), \\
\theta^4,\theta^8 & : & 6(0,0)_{-\frac{1}{3}}+6(0,0)_{-\frac{1}{6}}+
12(1,0)+12(0,1)+30(0,0), \\
\theta^6 & : & 6(0,0)_{-\frac{1}{4}}+6(1,0)+6(0,1)+12(0,0). 
\end{array}
\end{equation}
3$v$, 4$v$ and 6$v$ are one non-zero component vectors,
therefore the $\theta^3$+$\theta^9$, $\theta^4$+$\theta^8$ and $\theta^6$
sectors want some ($\frac{1}{2}$,$\frac{1}{2}$);
the $\theta^3$+$\theta^9$ sector wants 16($\frac{1}{2}$,$\frac{1}{2}$)
as compared with the $\theta$ sector of the $T^2$$\times$$T^2/Z_4$
orbifold with $v$=(0,0,$\frac{1}{4}$),
the $\theta^4$+$\theta^8$ sector does 24($\frac{1}{2}$,$\frac{1}{2}$)
as compared with the $\theta$ sector of the $T^2$$\times$$T^2/Z_3$
orbifold with $v$=(0,0,$\frac{1}{3}$),
and the $\theta^6$ sector does 12($\frac{1}{2}$,$\frac{1}{2}$)
as compared with the $\theta$ sector of the $T^2$$\times$$T^2/Z_2$
orbifold with $v$=(0,0,$\frac{1}{2}$). 
Each $\theta$ sector gives the standard number ratio of the states.  
We also find the tachyonic and massless spectrum of Type 0A string 
theory is
\begin{equation}
\begin{array}{lcl}
\theta^0 & : & (0,0)_{-\frac{1}{2}}+8(\frac{1}{2},\frac{1}{2})+4(0,0), \\
\theta,\theta^5,\theta^7,\theta^{11} & : & 24(0,0)_{-\frac{5}{24}}
+24(0,0)_{-\frac{1}{24}}+48(\frac{1}{2},\frac{1}{2}), \\
\theta^2,\theta^{10} & : & 
36(0,0)_{-\frac{1}{12}}+36(\frac{1}{2},\frac{1}{2}), \\
\theta^3,\theta^9 & : & 
4(0,0)_{-\frac{3}{8}}+8(0,0)_{-\frac{1}{8}}+16(\frac{1}{2},\frac{1}{2}), \\
\theta^4,\theta^8 & : &  
6(0,0)_{-\frac{1}{3}}+6(0,0)_{-\frac{1}{6}}
+24(\frac{1}{2},\frac{1}{2})+6(0,0), \\
\theta^6 & : & 6(0,0)_{-\frac{1}{4}}+12(\frac{1}{2},\frac{1}{2}). 
\end{array}
\end{equation}
Similar to Type 0B,
the $\theta^3$+$\theta^9$, $\theta^4$+$\theta^8$ and $\theta^6$ 
sectors want 8 DU, 12 DU and 6 DU as compared with the standard 
respectively. 

Since on this type $T^4/Z_N$ orbifolds
all of the states disappearing from the tachyonic and massless spectra 
come from R-R states,
we expect the corresponding fractional D-branes are also absent.

\section{Rules of the spectra}
\label{rule}

In this section we extract some rules of the
spectra of Type II/0 string theories on 
a $T^4/Z_N$ orbifold from the results in the previous section.
We focus on disappearing states and existence of tachyons 
in the twisted sectors of non-supersymmetric orbifolds.

First, we extract the rule of disappearing fermionic states from 
NS-R and R-NS sectors.
Since Type 0 string theories are bosonic, 
we focus only on Type II string theories.
There disappear some fermionic states in the $\theta^n$ sector
of non-supersymmetric $T^4/Z_N$ orbifolds, where $n$ is not relatively
prime to $N$.
However, there is an exceptional sector, that is, 
the $\theta^4$+$\theta^6$ sector of the $T^4/Z_{10}$ orbifold with 
$v$=(0,$\frac{1}{10}$,$\frac{3}{10}$).
It has no disappearing fermionic states.
 
Second, we consider disappearing bosonic states from NS-NS sectors
in not only Type II orbifolds but also Type 0 ones,
which have been treated in subsection \ref{t4.1},
where orbifolds do not include the $(-1)^{F_S}$=(0,0,1) twist.
Some NS-NS states disappear 
in the twisted sector with supersymmetric twist vector
of a non-supersymmetric $T^4/Z_N$ orbifold.
However, there is no disappearing NS-NS states
in the $\theta^2$+$\theta^{N-2}$ sector of a non-supersymmetric 
$T^4/Z_N$ orbifold even if the twist vector is supersymmetric.
We can take the $\theta^2$ sector of the $T^4/Z_4$ orbifold with  
$v$=(0,$\frac{1}{4}$,$\frac{3}{4}$) 
and the $\theta^2$+$\theta^4$ sector of 
that with $v$=(0,$\frac{1}{6}$,$\frac{5}{6}$) for instance.

Third, we extract the rule of disappearing states from R-R sectors, 
which we think is important for computing the D-brane spectrum.
In Type II and Type 0 string theories on a $T^4/Z_N$ orbifold, 
they disappear in the twisted sector whose twist vector has 
one non-zero component.
In this case disappearing does not depend on whether a $T^4/Z_N$ orbifold
includes the $(-1)^{F_S}$=(0,0,1) twist or not. 

Finally, we think about the relation between existence of tachyons
and the twist vector.
In a Type II orbifold, the tachyon appears in the twisted sector 
with non-supersymmetric twist vector. 
We can confirm this rule by
comparing the $\theta^3$+$\theta^9$ sectors of 
the two $T^4/Z_{12}$ orbifolds with each other:
one orbifold with $v$=(0,$\frac{1}{12}$,$\frac{5}{12}$) and the other
with $v$=(0,$\frac{1}{12}$,$\frac{7}{12}$).
Since the $\theta^3$+$\theta^9$ sector on the former 
has non-supersymmetric twist vector, there appear tachyons.
In contrast with that, the sector on the latter 
has supersymmetric twist vector, therefore there appear no tachyons. 
In a Type 0 orbifold,
there appear tachyons in any twisted sector except for in that
whose twist vector is (0,$\frac{1}{2}$,$\frac{1}{2}$),
where we take absolute value of two non-zero components.

\section{Summary and Discussions}

In this paper, 
we have found the orbifolds compatible with toroidal
compactification $T^2$ and $T^4$ and have enumerated the 
tachyonic and massless spectra of Type II and Type 0 string theories
on such a orbifold for advanced studies.
In Type 0 orbifolds,
we have obtained four pairs of those which could be identified:
(0,0,$\frac{1}{3}$)$\sim$(0,0,$\frac{2}{3}$),
(0,$\frac{1}{4}$,$-\frac{1}{4}$)$\sim$(0,$\frac{1}{4}$,$\frac{3}{4}$),
(0,$\frac{1}{3}$,$-\frac{1}{3}$)$\sim$(0,$\frac{1}{3}$,$\frac{2}{3}$),
and (0,$\frac{1}{5}$,$\frac{3}{5}$)$\sim$(0,$\frac{1}{5}$,$\frac{2}{5}$).
In addition, we extract some rules about disappearing states and
existence of tachyons.
Especially, disappearing states from R-R sectors seem to be 
the key to advanced studies.

On the basis of studies in this paper, we can expand.
For example, $Z_N$$\times$$Z_M$ orbifolds 
with and without torsion \cite{hep-th/0005153,fiqs,fiq,hep-th/0001200}, 
heterotic string orbifolds \cite{imnq,fiqs},
and Type II and Type 0 string theories on a $T^6/Z_N$ orbifold.
Part of the last example has studied in \cite{hepth0202057},
but there are many higher order non-supersymmetric $T^6/Z_N$ 
orbifolds not listed in it.

As another expansion along \cite{hepth9910109},
we can also consider the D-brane
spectra of Type II and Type 0 string theories  
on a $T^2/Z_N$ orbifold, and those on a $T^4/Z_N$. 
As mentioned above, we should remark the disappearing R-R states,
because we can expect corresponding fractional D-branes also disappear.
Therefore we should confirm that by computing the D-brane spectrum
with the boundary state formalism \cite{bsf}, 
and also with K-theory \cite{hepth9910109,hep-th/0005153,k-th,hep-th/0001200}.
It also seems worthwhile to expand a sequence of this studies into
$T^6/Z_N$ orbifolds and $Z_N$$\times$$Z_M$ orbifolds \cite{Z*Z}
in order to 
clarify the relation among the perturbative R-R states, the D-brane 
spectrum, and the K-theory.

\begin{center}
\textbf{Acknowledgements}
\end{center}
The author thanks K. Inoue for reading the manuscript and discussions.

\appendix
\section{Appendix : $T^2$$\times$$T^2/Z_N$ orbifolds}

In this appendix we transform the spectra we have enumerated in  
section \ref{t2} into those of $T^2$$\times$$T^2/Z_N$ version.
That is, those are written down as in section \ref{t4}.

The tachyonic and massless spectrum of Type IIB/IIA string theory on
the $T^2$$\times$$T^2/Z_3$ orbifold with $v$=(0,0,$\frac{2}{3}$) is
\begin{equation}
\begin{array}{lcl}
\theta^0 & : & 2(1,0)+2(0,1)+8(\frac{1}{2},\frac{1}{2})
+4(\frac{1}{2},0)+4(0,\frac{1}{2})+10(0,0), \\
\theta,\theta^2 & : & 6(0,0)_{-\frac{1}{3}}+6(1,0)+6(0,1)+
12(\frac{1}{2},\frac{1}{2})
+12(\frac{1}{2},0)+12(0,\frac{1}{2})+18(0,0).
\end{array}
\label{2/3II}
\end{equation}
And that of Type 0B/0A string theory is
\begin{equation}
\begin{array}{lcl}
\theta^0 & : & (0,0)_{-\frac{1}{2}}+
4(1,0)+4(0,1)+12(\frac{1}{2},\frac{1}{2})+14(0,0), \\
\theta,\theta^2 & : & 6(0,0)_{-\frac{1}{3}}+6(0,0)_{-\frac{1}{6}}
+12(1,0)+12(0,1)+24(\frac{1}{2},\frac{1}{2})+30(0,0).
\end{array}
\label{2/30}
\end{equation}
As done in section \ref{t2}, we regard the number ratio of the states
in each $\theta$+$\theta^2$ sector as the standard one with the
twist vector (0,0,$\frac{2}{3}$).
In Type 0 string theory, we regard the ratio as the standard one with 
the twist vector (0,0,$\frac{1}{3}$) also,
for we can identify the two $T^2$$\times$$T^2/Z_3$ orbifolds: 
one with $v$=(0,0,$\frac{1}{3}$) and the other 
with $v$=(0,0,$\frac{2}{3}$). 

The tachyonic and massless spectrum of Type 0B/0A string theory on
the $T^2$$\times$$T^2/Z_2$ orbifold with $v$=(0,0,$\frac{1}{2}$) is
\begin{equation}
\begin{array}{lcl}
\theta^0 & : & (0,0)_{-\frac{1}{2}}+
4(1,0)+4(0,1)+12(\frac{1}{2},\frac{1}{2})+16(0,0), \\
\theta & : & 8(0,0)_{-\frac{1}{4}}
+8(1,0)+8(0,1)+16(\frac{1}{2},\frac{1}{2})+16(0,0).
\end{array}
\end{equation}
We regard the number ratio of the states in the $\theta$ sector 
as the standard one with the twist vector (0,0,$\frac{1}{2}$).

The tachyonic and massless spectrum of Type 0B/0A string theory on
the $T^2$$\times$$T^2/Z_4$ orbifold with $v$=(0,0,$\frac{1}{4}$) is
\begin{equation}
\begin{array}{lcl}
\theta^0 & : & (0,0)_{-\frac{1}{2}}+
4(1,0)+4(0,1)+12(\frac{1}{2},\frac{1}{2})+14(0,0), \\
\theta,\theta^3 & : & 4(0,0)_{-\frac{3}{8}}+8(0,0)_{-\frac{1}{8}}
+8(1,0)+8(0,1)+16(\frac{1}{2},\frac{1}{2})+16(0,0), \\
\theta^2 & : & 6(0,0)_{-\frac{1}{4}}+
6(1,0)+6(0,1)+12(\frac{1}{2},\frac{1}{2})+12(0,0). 
\end{array}
\end{equation}
We also regard the number ratio of the states in the $\theta$+$\theta^3$
sector as standard one with the twist vector (0,0,$\frac{1}{4}$).

The tachyonic and massless spectrum of Type 0B/0A string theory on
the $T^2$$\times$$T^2/Z_6$ orbifold with $v$=(0,0,$\frac{1}{6}$) is
\begin{equation}
\begin{array}{lcl}
\theta^0 & : & (0,0)_{-\frac{1}{2}}+
4(1,0)+4(0,1)+12(\frac{1}{2},\frac{1}{2})+14(0,0), \\
\theta,\theta^5 & : & 2(0,0)_{-\frac{5}{12}}+2(0,0)_{-\frac{1}{4}}+
4(0,0)_{-\frac{1}{12}}
+4(1,0)+4(0,1)+8(\frac{1}{2},\frac{1}{2})+8(0,0), \\
\theta^2,\theta^4 & : & 4(0,0)_{-\frac{1}{3}}+4(0,0)_{-\frac{1}{6}}+
8(1,0)+8(0,1)+16(\frac{1}{2},\frac{1}{2})+20(0,0), \\
\theta^3 & : & 4(0,0)_{-\frac{1}{4}}+
4(1,0)+4(0,1)+8(\frac{1}{2},\frac{1}{2})+8(0,0). 
\end{array}
\end{equation}
We regard the number ratio of the states in the $\theta$+$\theta^5$
sector as standard one with the twist vector (0,0,$\frac{1}{6}$).

Through this appendix,
we can confirm that, on $T^2$$\times$$T^2/Z_N$ orbifolds,
each number ratio of the states is the same 
as long as twisted sectors have the same twist vector.



\begin{thebibliography}{99}
\bibitem{nso}
R. Blumenhagen, A. Font and D. Lust,
\textit{Non-Supersymmetric Gauge Theories from D-Branes in Type 0 String Theory},
Nucl. Phys. B560 (1999) 66, hep-th/9906101;\\
%
I. R. Klebanov, N. A. Nekrasov and S. L. Shatashvili,
\textit{An Orbifold of Type 0B Strings and Non-supersymmetric Gauge Theories},
Nucl. Phys. B591 (2000) 26, hep-th/9909109;\\
%
O. Bergman and M. R. Gaberdiel,
\textit{On the Consistency of Orbifolds},
Phys. Lett. B481 (2000) 379, hep-th/0001130.
\bibitem{dhsw}
L. J. Dixon and J. A. Harvey,
\textit{String theories in ten dimensions without spacetime supersymmetry}, 
Nucl. Phys. B274 (1986) 93;\\
N. Seiberg and E. Witten,
\textit{Spin structures in string theory},
Nucl. Phys. B276 (1986) 272.
\bibitem{dhvw}
L. Dixon, J. Harvey. C. Vafa and E. Witten,
\textit{Strings on Orbifolds}, Nucl. Phys. B261 (1985) 678;
\textit{ Strings on Orbifolds II}, Nucl. Phys. B274 (1986) 285. 
\bibitem{hepth9910109}
M. R. Gaberdiel and B. Stefanski Jr,
\textit{Dirichlet Branes on Orbifolds},
Nucl. Phys. B578 (2000) 58, hep-th/9910109.
\bibitem{hep-th/0005153}
B. Stefanski Jr,
\textit{Dirichlet Branes on a Calabi-Yau Three-fold Orbifold},
Nucl. Phys. B589 (2000) 292, hep-th/0005153
\bibitem{hepth0202057}
A. Font and A. Hernandez,
\textit{Non-Supersymmetric Orbifolds}, 
Nucl. Phys. B634 (2002) 51, hep-th/0202057.
\bibitem{pol}
J. Polchinski,
\textit{String Theory: Superstring Theory and Beyond}, 
Cambridge University Press, 1998.
\bibitem{tac}
A. Adams, J. Polchinski and E. Silverstein,
\textit{Don't Panic! Closed String Tachyons in ALE Spacetimes},
JHEP 0110 (2001) 029, hep-th/0108075;\\
%
C. Vafa,
\textit{Mirror Symmetry and Closed String Tachyon Condensation},
hep-th/0111051;\\
%
Y. Michishita and P. Yi,
\textit{D-Brane Probe and Closed String Tachyons},
Phys. Rev. D65 (2002) 086006, hep-th/0111199;\\
%
J. R. David, M. Gutperle, M. Headrick and S. Minwalla,
\textit{Closed String Tachyon Condensation on Twisted Circles},
JHEP 0202 (2002) 041, hep-th/0111212;\\ 
%
S.P. de Alwis and A.T. Flournoy,
\textit{Closed String Tachyons and Semi-Classical Instabilities},
Phys.Rev. D66 (2002) 026005, hep-th/0201185;\\
%
S. Minwalla and T. Takayanagi,
\textit{Evolution of D-branes Under Closed String Tachyon Condensation},
JHEP 0309 (2003) 011, hep-th/0307248;\\
%
T. Suyama,
\textit{On decay of Bulk Tachyons},
hep-th/0308030;\\
%
Y. Imamura,
\textit{Decay of type 0 NS5-branes to nothing},
Phys. Rev. D69 (2004) 026005, hep-th/0309024.
\bibitem{hep-th/9906055}
O. Bergman and M.R. Gaberdiel,
\textit{Dualities of Type 0 Strings},
JHEP 9907 (1999) 022, hep-th/9906055.
\bibitem{hepth9207111}
J. Erler and A. Klemm,
\textit{Comment on the Generation Number in Orbifold Compactifications},
Commun. Math. Phys. 153 (1993) 579, hep-th/9207111.
\bibitem{imnq}
L. E. Ib\' a\~ nez, J. Mas, H. P. Nilles and F. Quevedo,
\textit{Heterotic strings in symmetric and asymmetric orbifold backgrounds}, 
Nucl. Phys. B301 (1988) 157. 
\bibitem{fiqs}
A. Font, L. E. Ib\' a\~ nez, F. Quevedo and A. Sierra,
\textit{The construction of "realistic" four-dimensional strings
through orbifolds},
Nucl. Phys. B331 (1990) 421.
\bibitem{fiq}
A. Font, L. E. Ib\' a\~ nez and F. Quevedo,
\textit{$Z_N \times Z_M$ orbifolds and discrete torsion},
Phys. Lett. B 217 (1989) 272. 
\bibitem{bsf}
D. Diaconescu and J. Gomis,
\textit{Fractional Branes and Boundary States in Orbifold Theories},
JHEP 0010 (2000) 001, hep-th/9906242;\\
%
M. R. Gaberdiel,
\textit{Lectures on Non-BPS Dirichlet branes},
Class. Quant. Grav. 17 (2000) 3483.
\bibitem{k-th}
E. Witten,
\textit{D-Branes And K-Theory},
JHEP 9812 (1998) 019, hep-th/9810188;\\
%
H. Garcia-Compean,
\textit{D-branes in Orbifold Singularities and Equivariant K-Theory},
Nucl. Phys. B557 (1999) 480, hep-th/9812226;\\
%
K. Olsen and R. J. Szabo,
\textit{Constructing D-Branes from K-Theory},
Adv. Theor. Math. Phys. 3 (1999) 889, hep-th/9907140;\\
%
E. J. Martinec and G. Moore,
\textit{On Decay of K-theory},
hep-th/0212059.
\bibitem{hep-th/0001200}
J. Gomis,
\textit{D-branes on Orbifolds with Discrete Torsion And Topological Obstruction},
JHEP 0005 (2000) 006, hep-th/0001200.
\bibitem{Z*Z}
C. Vafa and E. Witten,
\textit{On Orbifolds with Discrete Torsion},
J. Geom. Phys. 15 (1995) 189, hep-th/9409188;\\
%
M. R. Douglas,
\textit{D-branes and Discrete Torsion},
hep-th/9807235;
\textit{D-branes and Discrete Torsion II},
hep-th/9903031;\\
%
D. Berenstein, R. G. Leigh,
\textit{Discrete Torsion, AdS/CFT and duality},
JHEP 0001 (2000) 038, hep-th/0001055;\\
%
M. R. Gaberdiel,
\textit{Discrete torsion orbifolds and D-branes},
JHEP 0011 (2000) 026, hep-th/0008230;\\
B. Craps and M. R. Gaberdiel,
\textit{Discrete torsion orbifolds and D-branes II},
JHEP 0104 (2001) 013, hep-th/0101143. 

\end{thebibliography}
\end{document}